\documentstyle[12pt]{article}
\newcommand{\om}{\omega}
\newcommand{\al}{\alpha}
\newcommand{\ep}{\epsilon}

\newcommand{\deebar}{\bar{\partial}}

\newcommand{\msc}[1]{\mbox{\scriptsize #1}}
\newcommand{\dsp}{\displaystyle}

\newcommand{\bc}{\mbox{{\bf C}}}
\newcommand{\br}{\mbox{{\bf R}}}
\newcommand{\bz}{\mbox{{\bf Z}}}

\newcommand{\bsz}{\msc{{\bf Z}}}

\newcommand{\ket}[1]{{|#1\rangle}}
\newcommand{\bra}[1]{{\langle#1|}}

\newcommand{\mod}{\mbox{mod}}

\newcommand{\cL}{{\cal L}}
\newcommand{\cJ}{{\cal J}}
\newcommand{\cG}{{\cal G}}
\newcommand{\cA}{{\cal A}}
\newcommand{\cF}{{\cal F}}
\newcommand{\cN}{{\cal N}}
\newcommand{\cD}{{\cal D}}
\newcommand{\cE}{{\cal E}}
\newcommand{\cO}{{\cal O}}
\newcommand{\cW}{{\cal W}}
\newcommand{\cH}{{\cal H}}

\newcommand{\lsim}{\stackrel{<}{\sim}}

\newcommand {\eqn}[1]{(\ref{#1})}

\makeatletter
\@addtoreset{equation}{section}

\makeatother

\begin{document}
\vskip 5mm
%%% Title page %%%%%
\begin{titlepage}
 
 \renewcommand{\thefootnote}{\fnsymbol{footnote}}
 \font\csc=cmcsc10 scaled\magstep1
 {\baselineskip=15pt
 \rightline{
 \vbox{\hbox{hep-th/0005065}
       \hbox{UT-887}
       \hbox{YITP-00-22}      
 }}}

 \vfill
 \baselineskip=20pt
 \begin{center}
 \centerline{\Huge  String Theory on $AdS_3$} 
 \vskip 5mm 
 \centerline{\Huge as }
  \vskip 5mm 
 \centerline{\Huge Discrete Light-Cone Liouville Theory}

 \vskip 0.7 truecm

\noindent{\it \large Yasuaki
  Hikida${}^a$\footnote{hikida@hep-th.phys.s.u-tokyo.ac.jp}, 
  Kazuo Hosomichi${}^b$\footnote{hosomiti@yukawa.kyoto-u.ac.jp} 
 and Yuji Sugawara${}^c$\footnote{sugawara@hep-th.phys.s.u-tokyo.ac.jp}} \\

\bigskip
 \vskip .4 truecm
 {\baselineskip=10pt
${}^{a,c}$\hspace{-.1cm} {\it Department of Physics,  Faculty of Science, 
  University of Tokyo \\
  Bunkyo-ku, Hongo 7-3-1, Tokyo 113-0033, Japan}
 } \\
 \vskip .2 truecm
${}^b$\hspace{-.1cm} {\it Yukawa Institute for Theoretical Physics, Kyoto
  University,\\  Kyoto 606-8502, Japan}

 \vskip .4 truecm

 \end{center}

 \vfill
 \vskip 0.5 truecm

\begin{abstract}
\baselineskip 6mm

We investigate  (super) string theory on $AdS_3$ background 
based on an approach of free field realization. 
We demonstrate that this string theory can be reformulated 
as a string theory defined on a linear dilaton background 
along the transverse direction (``Liouville mode'')
and compactified onto $S^1$ along a {\em light-like\/} direction.

Under this reformulation we analyze 
the physical spectrum as that of a free field system, 
and discuss the consequences when we turn on the Liouville 
potential. The discrete light-cone momentum in our framework
is naturally interpreted  as the ``winding number'' of the long string 
configuration and is identified with the spectral flow 
parameter that is introduced in the recent work by Maldacena and Ooguri 
\cite{MO}. 

Moreover we show that there exist infinite number of the on-shell chiral 
primary states possessing the different light-cone momenta 
and the spectral flow consistently acts on the space of chiral primaries.
We observe that they are also chiral primaries in the sense of
space-time (or the conformal theory of long string) and the spectrum
of space-time $U(1)_R$ charge is consistent with the expectation from
the $AdS_3/CFT_2$-duality. We also clarify the correspondence between
our framework and the symmetric orbifold theory 
of multiple long string system \cite{HS2}.

\end{abstract}

\setcounter{footnote}{0}
\renewcommand{\thefootnote}{\arabic{footnote}}
\end{titlepage}

\newpage
\baselineskip 7mm

%%%%%%%%%%%%%%%%%%%%%%%%%%%%%%%%%%%%%%%%%%%%%%%%%%%%%%%%%%%%%%%%%%%%%%%%%%
\section{Introduction}

\hspace*{4.5mm}

Study of string theory on $AdS_3$ background has been a subject
of great importance mainly for the following two reasons: Firstly, it is
a non-trivial example of completely solvable string theories on 
a curved background with the Lorentzian signature \cite{sl2,Bars,Satoh}. 
The second reason, which is comparably newer than the first one, 
is the possibility of understanding the $AdS/CFT$-duality \cite{AdS}
at a stringy level \cite{GKS,DORT,KS}.  

Although the string theory on this background (with NS $B$-field)
is believed  to be described by a simple conformal field theory, 
$SL(2;\br)$-WZW model\cite{GKS}, there are still subtleties.
Especially two different theoretical grounds have been 
proposed with respect to the set up of the Hilbert space of quantum states;
\begin{enumerate}
 \item The Hilbert space is defined to be a representation space of 
        current algebra $\widehat{SL}(2;\br)$.
 \item The Hilbert space is defined to be the Fock space of free
       fields which should be identified with the string coordinates 
        in some suitable parametrizations of the group manifold $SL(2;\br)$. 
\end{enumerate}
The first ground is based on the standard  prescription of two
dimensional conformal field theory ($CFT_2$). It is well-known that
the WZW model for a compact group actually has such a Hilbert space:
the physical Hilbert space should be made up of a finite number of the 
unitary (integrable) representations of current algebra.
Since we now have a non-compact group $SL(2;\br)$, the situation becomes
rather non-trivial. If $k\equiv k'-2$ ($k'$ means the level 
of $\widehat{SL}(2;\br)$) is equal to a negative rational value, 
we have a finite number of the ``admissible'' representations\footnote
  {The admissible representations are not necessarily unitary
   representations of $\widehat{SL}(2;\br)$ with $ k=-q/p$. 
   But these define the good-natured  conformal blocks 
   in the similar manner as the familiar  $(p,q)$ minimal model with 
   $\dsp c=1-\frac{6(p-q)^2}{pq}$ \cite{BF}.}  which 
contain  a  rich structure with many singular vectors \cite{BF}. 
However, string theory on $AdS_3$ background corresponds to 
the cases of {\em positive\/} $k$, in which  we cannot have  
any unitary representation of $\widehat{SL}(2;\br)$ 
(except for the trivial representation). 
There are infinite number of non-unitary representations
some of which have a few singular vectors 
(generically, no singular vectors).
The best we can expect is that the BRST condition 
successfully eliminates all negative norm states from 
the physical Hilbert space.
Many discussions have been given concerning the no-ghost theorem 
along this line \cite{sl2}. With respect to the discrete series
the no-ghost theorem was proved under the assumption of a truncation of 
quantum number ``$j$''  parametrizing the second Casimir \cite{EGP}.

In this traditional approach to $CFT_2$ from the representation 
theory of current algebra, primary states 
are naively  characterized by the two quantum numbers $j$, $m$.
However, as was carefully discussed by Bars \cite{Bars}, 
there is a subtle point if we recall that 
we are working in a $\sigma$-model 
{\em with a three dimensional non-compact target space.}  
One should keep it in mind that 
under the limit of weak curvature; 
$k\longrightarrow +\infty$, $AdS_3$ space (the universal covering 
of $AdS_3$, strictly speaking) may  be replaced by a flat background 
$\br^{2,1}$. In this sense it may be more natural 
that we have {\em three\/} conserved momenta characterizing 
the physical states, and the second definition of Hilbert space 
is likely  to be more appropriate, as was claimed in
the works \cite{Bars,Satoh,BDM}.

More recently, based on the representation theory of $\widehat{SL}(2;\br)$,
Maldacena and Ooguri claimed \cite{MO} that one should take 
the Hilbert space enlarged by the spectral flow and proved
the extended no-ghost theorem. 
On physical ground it may be plausible that 
the third momentum we mentioned above is related to 
such an enlargement of Hilbert space. 
One of the main purposes of this paper is to clarify the relation
between them, namely, the role of spectral flow in the argument of \cite{MO} 
and that of the third momentum of zero-modes \cite{Bars,Satoh,BDM} 
(discrete light-cone momentum in the context of this paper).

Another aim of this work is to manifest  further the role of
``space-time Virasoro  algebra'' introduced in \cite{GKS}.
It is inspired by the asymptotic isometries of Brown-Henneaux
\cite{BH,Strominger} and is understood   to describe the conformal 
symmetry of the long string sector \cite{MMS,SW}.
The generators of this algebra are most conveniently
realized as operators acting on the Fock space 
of free fields (the Wakimoto's  $\varphi$, $\beta$, 
$\gamma$ system in the usual treatment). 
This is one of the reasons why we take the free field 
realization rather than the abstract representation theory of 
current algebra.

%We make more one comment about the no-ghost theorem in 
%the discrete series. The truncation of $j$ physically 
%means that we have an upper bound in the string mass spectrum.
%This statement sounds artificial and unnatural,
%because  one expect that usual string spectrum includes large number of 
%heavy Kaluza-Klein or winding excitations with 
%no upper bounds of mass-squared.
%Recently,  Maldacena and Ooguri \cite{MO} discussed how one can resolve this 
%difficulty based on the representation theory of $\widehat{SL}(2;\br)$. 
%They showed  that the enhancement  of the representation space 
%by the spectral flow leads to an extended no-ghost theorem compatible with 
%the mass spectrum {\em with no upper bound.} 
%One of the main purposes of this paper is to clarify the relation
%between the role of spectral flow in the argument of \cite{MO} 
%and that of the  third momentum of zero-modes 
%(discrete light-cone momentum in the context of this paper) 
%which is emphasized in \cite{Bars,Satoh,BDM} and mentioned above.

~

This paper is organized as follows;

In section 2 we start with reformulating the bosonic and superstring 
theories on $AdS_3$ background by a free field realization. 
With the help of some field redefinitions we show that
the $AdS_3$ string theory can be described by 
a string theory on a  linear dilaton background 
(along the transverse direction) and with a 
{\em light-like compactification\/}.
We further demonstrate that the space-time conformal algebra 
given in  \cite{GKS} has a quite simple form 
analogous to the DDF operators \cite{DDF}
in this framework of ``discrete light-cone Liouville theory''.

In section 3 we analyze the physical spectrum in our framework.
As that of a free field system we will reproduce  the spectrum proposed 
in \cite{Bars,Satoh,BDM}: Only the principal series is allowed 
due to the unitarity. We also comment  on the outcome of 
turning on the Liouville potential term,
which should be a marginal perturbation. 
Such an interaction term breaks the translational invariance along 
the radial direction (in other words, makes the ``screening'' of the extra
zero-mode momentum from the view points of the $SL(2;\br)$ current
algebra), and hence the  ``bound string states'' possessing the imaginary 
momentum along this direction may appear in the physical spectrum. 
The space-time Virasoro operators play a role as  
the {\em spectrum generating algebra\/}, and we will observe
that one must include the representations of $\widehat{SL}(2;\br)$ 
which are broader than those given in \cite{MO} in order to   
make the full space-time Virasoro algebra act successfully on 
the physical Hilbert space.

In section 4 we further investigate the spectrum for superstring cases
that give rise to space-time $N=2, 4$ SCFTs. We present 
the complete set of on-shell chiral primaries.
We will find  that there are infinite number of on-shell 
chiral primary states with the  different light-cone momenta,
and the spectral flows act naturally among them. They become the chiral 
primaries also in the sense of {\em space-time}
and have the space-time $U(1)_R$ charges in agreement  with 
the expectation of $AdS_3/CFT_2$-duality.
As a byproduct we also clarify the relationship with 
the symmetric orbifold CFT describing the multiple long strings 
discussed in \cite{HS2}.

We will summarize the main results of our analyses and present some
discussions in section 5.

~

\section{Reformulation of $AdS_3$ String Theory as the Discrete
Light-cone Liouville Theory}

\hspace*{4.5mm}

\subsection{Bosonic String on $AdS_3 \times { \cal N }$}

\hspace*{4.5mm}

Through this paper we shall consider the universal covering
of the $AdS_3$ space with the Lorentzian signature so that 
the time direction is non-compact.
We start with the following world-sheet Lagrangian for 
the (quantum) bosonic string on $AdS_3$ with $NS$ B-filed \cite{GKS}
\begin{equation}
\cL=\partial\varphi\deebar\varphi- \sqrt{\frac{2}{k}}R^{(2)}\varphi
+\beta\deebar\gamma+\bar{\beta}\partial\bar{\gamma}
-\beta\bar{\beta}\exp \left(-\sqrt{\frac{2}{k}} \varphi\right),
\label{ads3 lagrangian}
\end{equation} 
where $R^{(2)}$ denotes the curvature on the world-sheet.
Throughout this paper we shall only focus on 
the physics at the near boundary region 
$\varphi \sim +\infty$, in which we can consider 
the interaction term (``screening charge term'') 
$\dsp  \beta\bar{\beta}\exp \left(-\sqrt{\frac{2}{k}}\varphi\right)$
as a small perturbation.
By dropping this term simply
we obtain the free conformal field theory 
\begin{equation}
T=-\frac{1}{2}\partial \varphi \partial \varphi 
-\frac{1}{\sqrt{2k}}\partial^2\varphi +\beta\partial \gamma,
\end{equation}
\begin{equation}
 \varphi(z)\varphi(0) \sim - \ln z~,~~ \beta(z)\gamma(0) \sim \frac{1}{z}~.
\end{equation}
We will later discuss the effect of restoring this interaction term 
$\dsp \beta\bar{\beta}\exp \left(-\sqrt{\frac{2}{k}}\varphi\right)$ 
on the physical spectrum.

The $SL(2;\br)$ symmetry in this free system is described by 
the Wakimoto representation \cite{Wakimoto}
\begin{eqnarray}
  j^- &=& \beta  \nonumber \\
  j^3 &=& \beta\gamma +  \sqrt{\frac{k}{2}}\partial\varphi  \label{wakimoto}\\
  j^+ &=& \beta\gamma^2 + \sqrt{2k}\gamma\partial\varphi +
  (k+2)\partial\gamma~, \nonumber 
\end{eqnarray}
which generates the $\widehat{SL}(2;\br)$ current algebra of level $k+2$
\begin{equation}
\left\{
\begin{array}{lll}
j^3(z)j^3(0) &\sim& \dsp -\frac{(k+2)/2}{z^2} \\
j^3(z)j^{\pm}(0)&\sim& \dsp \frac{\pm 1}{z}j^{\pm}(0) \\
j^+(z)j^-(0) &\sim& \dsp \frac{k+2}{z^2}-\frac{2}{z}j^3(0) ~.
\end{array}
\right. 
\end{equation}

By using the standard  bosonization of  $\beta,\gamma$ \cite{FMS} 
\begin{equation}
 \beta = i \partial v e^{-u -iv}~,~~ \gamma = e^{u+iv}~, 
~~~ u(z)u(0)\sim -\ln z, ~v(z)v(0)\sim -\ln z,
\end{equation}
we can rewrite the currents  \eqn{wakimoto}
\begin{eqnarray}
  j^- &=& i \partial v e^{-u -iv}  \nonumber \\
  j^3 &=& -\partial u + \sqrt{\frac{k}{2}}\partial \varphi \\
  j^+ &=& e^{u+iv} 
       (k \partial (u+iv)+\sqrt{2k}\partial \varphi + i \partial v) 
 \nonumber~~.
\end{eqnarray}
Moreover it is convenient to introduce the following  new variables;
\begin{eqnarray}
Y^0 &:=&  \sqrt{\frac{2}{k+2}} iu -\sqrt{\frac{k}{k+2}} i \varphi \nonumber\\
Y^1 &:=& - \sqrt{\frac{k+2}{2}}v + \frac{k}{\sqrt{2(k+2)}}iu 
    + \sqrt{\frac{k}{k+2}} i \varphi \label{YYrho}\\
\rho &:=&  \sqrt{\frac{k}{2}} (u +iv) + \varphi \nonumber~.
\end{eqnarray}
Since this field redefinition is an $SO(2,1)$-rotation, we 
simply have
\begin{equation}
 Y^0(z)Y^0(0) \sim \ln z ~,~~ Y^1(z) Y^1(0) \sim - \ln z ~,~~ 
 \rho (z) \rho (0) \sim - \ln z~~,
\label{OPE1}
\end{equation}
and any other combinations have no singular OPEs.

In these variables the $\widehat{SL}(2;\br)_{k+2}$ currents are given by
\begin{eqnarray}
  j^3 &=& \sqrt{\frac{k+2}{2}}i \partial Y^0  \nonumber \\
  j^{\pm} &=& \left( - \sqrt{\frac{k+2}{2}} i \partial Y^1  \pm 
     \sqrt{\frac{k}{2}}  \partial \rho  \right) 
       e^{\mp  \sqrt{\frac{2}{k+2}} i(Y^0 + Y^1) }~~,
\label{sl2}
\end{eqnarray}
and also the stress tensor is rewritten as 
\begin{equation}
 T = \frac{1}{2}(\partial Y^0)^2 - \frac{1}{2} (\partial Y^1)^2
 - \frac{1}{2} (\partial \rho)^2 - \frac{1}{\sqrt{2k}} \partial^2 \rho~,
\end{equation}
which of course has the correct central charge $\dsp c=3+\frac{6}{k}$.

In this way we have found that 
the bosonic string theory on $AdS_3$ can be realized by 
two free bosons $Y^0$, $Y^1$ (with no background charge) 
and a ``Liouville mode'' $\rho$
with the background charge $\dsp Q\equiv\sqrt{\frac{2}{k}}$. 
The essentially same realizations of $\widehat{SL}(2;\br)$
were used in several works \cite{IK,Satoh,YZ}. 
It was suggested in \cite{Mizoguchi} that the fields
$Y^0,~Y^1,~\rho$ roughly corresponds to the global coordinates of 
$AdS_3$ space, 
and in the similar sense  one might suppose that the Wakimoto fields 
$\varphi,~\beta,~\gamma$ correspond to
the Poincar\'{e} coordinates.

The definitions of  Hermitian conjugations  of these free fields  
are standard
\begin{equation}
(Y^i(z))^{\dag} = Y^i(1/z),~~~ (\rho(z))^{\dag} = \rho(1/z)-Q\ln z.
\label{hc}
\end{equation}
(Recall that the Liouville field $\rho$ has a  background charge.)
One can easily verify  that the Hermitian conjugations  of the current
generators take the usual forms  $j_n^{3\dag}=j^3_{-n}$, 
$j^{\pm\dag}_n=j^{\mp}_{-n}$ under \eqn{hc}.
This is an advantage of this free field realization \eqn{sl2} compared with 
the Wakimoto representation in which the rules of Hermitian conjugation 
are not simple. 

Notice also that the OPE of $Y^0 $ with itself has the wrong sign \eqn{OPE1}
and thus $Y^0$ should correspond to the time-like coordinate. 
This fact is consistent with the usual interpretation;
$j^3_0 \sim \mbox{energy}$.

There is a subtle point with respect to the realizations of currents
\eqn{sl2}: $j^{\pm}(z)$ are not necessarily defined as local operators 
on the whole Fock space of $Y^0$, $Y^1$, $\rho$. 
To overcome this difficulty it is natural to assume the following 
light-like compactification.
Let us introduce the light-cone coordinates
\begin{equation}
 Y^{\pm} = \frac{1}{\sqrt{2}}(Y^0 \pm Y^1)~,
\end{equation}
and assume the periodic identification
\begin{equation}
 Y^- \sim Y^- + 2 \pi R~,~~R = \frac{2}{\sqrt{k+2}}~~.
\label{llc}
\end{equation}
Such a prescription is known as 
the name  ``discrete light-cone quantization'',
and have been applied to the studies of M(atrix) theory with finite $N$
\cite{DLCQ}.

In this case, the conjugate momentum of $Y^-$ is quantized as 
\begin{equation}
 \frac{\partial Y^+}{\partial \tau} \equiv P^+ + \bar{P}^+ = \frac{2n}{R}~, 
 ~~n \in {\bf Z}~,
\end{equation}
and the winding mode of $Y^-$ should take
\begin{equation}
  \frac{\partial Y^-}{\partial \sigma} \equiv  P^- -  \bar{P}^- = m R~,
 ~~m \in {\bf Z}~~.
\label{winding Y-}
\end{equation} 
Since $Y^+$ remains  non-compact, there is no winding
mode along this direction
\begin{equation}
   \frac{\partial Y^+}{\partial \sigma} \equiv  P^+ -  \bar{P}^+ = 0~,
\end{equation}
and thus we obtain 
$\dsp P^+ = \bar{P}^+ = \frac{n}{R}$ $\dsp (n\in \bz)$.

By using these facts, we can concretely
write the ``tachyon'' vertex operators 
${\cal V}_{j,m,\bar{m},p}(z,\bar{z})\equiv 
V_{j,m,p}(z)\bar{V}_{j,\bar{m},p}(\bar{z})$,
where the left mover is defined by
\begin{equation}
 V_{j,m,p} = e^{ \left( \frac{\sqrt{k+2}}{2} p 
 - \frac{2m}{\sqrt{k+2}} \right) i Y^+ 
 + \frac{\sqrt{k+2}}{2}p i Y^- - \sqrt{\frac{2}{k}}j \rho}~~,
\label{vertex1}
\end{equation}
and the winding condition \eqn{winding Y-} means that $m-\bar{m}\in\bz$.
The corresponding Fock vacuum $\ket{j,m,p}$ has the following properties;
\begin{eqnarray}
j^3_0\ket{j,m,p}&=&(m-\frac{k+2}{2}p)\ket{j,m,p}, \\
j^{\pm}_{\mp p}\ket{j,m,p}&=&(m\pm j)\ket{j,m\pm 1, p}, \\
j^{\pm}_{\mp p+n}\ket{j,m,p}&=&0 , ~~~(\forall n \geq 1), \\
L_0\ket{j,m,p} &=& 
\left(-\frac{1}{k}j(j-1)+mp-\frac{k+2}{4}p^2\right)\ket{j,m,p} .
\end{eqnarray}
Namely, $j$, $m$ mean the quantum numbers appearing in the usual  
$\widehat{SL}(2;\br)$ theory and $p$ corresponds
to the label of ``flowed representation'' of \cite{MO}. 
In fact,  the spectral flow in the context of \cite{MO} is defined 
as the following transformations  
\begin{equation}
\left\{
\begin{array}{l}
\dsp  j^3(z) ~ \longrightarrow ~ j^3(z) + \frac{k+2}{2} \frac{p}{z}\\
j^{\pm}(z) ~\longrightarrow ~ z^{\mp p}\, j^{\pm}(z) ~.
\end{array}
\right.
\label{sflow1}
\end{equation}
In the system of $Y^0$, $Y^1$, $\rho$, this is simply  
the momentum shift 
\begin{equation}
\dsp Y^0~\longrightarrow~Y^0
-p\sqrt{\frac{k+2}{2}}i \ln z, 
\label{sflow2}
\end{equation}
and $Y^1$, $\rho$ remain unchanged.

The global $SL(2;\br)$ algebra $\{j^3_0,~j^{\pm}_0\}$ is manifestly
BRST invariant. We can  immediately  extend this algebra to 
the ``space-time Virasoro algebra''
\begin{equation}
\left\{
\begin{array}{lll}
 \cL_0 &=& \dsp -j^3_0 \equiv - \oint \sqrt{\frac{k+2}{2}}i\partial Y^0 \\
 \cL_n &=&\dsp  \oint \left( \sqrt{\frac{k+2}{2}}  
 i\partial Y^1 - n \sqrt{\frac{k}{2}}
 \partial \rho\right) e^{- \frac{2n}{\sqrt{k+2}} iY^+}~~~(n\neq 0),
\end{array}
\right.
\label{stV1}
\end{equation}
which actually generates the Virasoro algebra on the Fock space
over the vacuum $\ket{j,m,p}$
\begin{equation}
 [{\cal L}_n,{\cal L}_m] = (n-m){\cal L}_{n+m} + 
\frac{c}{12}(n^3-n) \delta_{n+m,0} ~,
\end{equation}
where $c=6(k+2)p$. 

We here make a few comments:
Firstly, the Virasoro operators $\cL_n$ are well-defined 
as local operators on the whole Fock  space compatible with the 
light-like compactification \eqn{llc}.  
Secondly, it is natural to regard   $\cL_n$ ($n\neq 0$) 
as  analogs  of the DDF operators \cite{DDF} along the $\rho$-direction.
In fact, $\cL_n$ ($n\neq 0$) is no other than the unique solution
for the BRST condition among the operators having the form 
$\dsp \sim \oint (A\partial \rho +B\partial Y^+ +C\partial Y^-) 
e^{-\frac{2n}{\sqrt{k+2}}iY^+}, ~A\neq 0 $
(up to BRST exact terms and some overall constant, of course).
In the next section we will make use of such  DDF like operators
in order to construct the complete set of the physical states. 
As the last comment, we should  point out that
$\cL_n$  are BRST equivalent to the space-time Virasoro operators
constructed in \cite{GKS}. It is straightforward to confirm that 
the quantum number $p$ precisely coincides with the ``winding of $\gamma$''; 
$\dsp \oint \gamma^{-1}d\gamma =p$.

%%%%%%%%%%%%%%%%%%%%%%%%%%%%%%%%%%%%%%%%%%%%%%%%%%%%%%%%%%%%%%%%%%%%

~

\subsection{Superstring on $AdS_3 \times S^1 \times { \cal N }/U(1)$}

\hspace*{4.5mm}

Let us try to extend our  previous results  
to the superstring cases.
We start with the general superstring vacua 
$AdS_3 \times S^1 \times { \cal N }/U(1)$ studied  in \cite{GRBL},
which are compatible with the world-sheet $N=2$ SUSY. 
The most familiar example $AdS_3 \times S^3 \times M^4$
($M^4=T^4$ or $K3$) is nothing but a special example of these
backgrounds, and we can readily apply the results 
in this subsection to that case too. 

First of all, to fix the notations 
we summarize the world-sheet properties of this superstring model;
\begin{itemize}
 \item $AdS_3$ sector ($ j^A,\psi ^A $)\\
To extend to the superstring case, we introduce free fermions in
the adjoint representation
\begin{eqnarray}
 \psi^3(z) \psi^3(0) \sim - \frac{1}{z}&,&
 \psi^+(z) \psi^-(0) \sim \frac{2}{z}~,\nonumber\\
\psi^{\pm} &=& \psi^1 \pm i \psi^2~~.
\end{eqnarray}
The total $\widehat{SL}(2;\br)$ currents are given by 
 \begin{equation}
  J^A = j^A + j^A_F = j^A - \frac{i}{2}\epsilon^A_{~BC}\psi ^B\psi^C~,
  ~~~~A,B,C=1,2,3~,
 \end{equation}
where the fermionic currents $j^A_F$ have the level $-2$. 
The fermionic currents $j^A_F$ can be written by free fermions as
\begin{equation}
 j_F^{\pm} = \pm \psi^{\pm}\psi^3~,
~~j^3_F = \frac{1}{2}\psi^+ \psi^-~~. 
\end{equation}

This sector has an $N = 1$ superconformal symmetry given by
\begin{eqnarray}
 G_{SL(2,{\bf R})} &=& 
\sqrt{\frac{2}{k}}(\frac{1}{2}\psi^+ j^- 
 + \frac{1}{2}\psi^- j^+ - \psi^3 j^3
 - \frac{1}{2} \psi^+ \psi^- \psi^3)~,
\end{eqnarray}
and the central charge is 
\begin{equation}
 c = \frac{3(k+2)}{k} + \frac{3}{2}~~.
\end{equation}
\end{itemize}

\begin{itemize}
 \item $S^1$ sector ($Y,\chi$)\\
We have a scalar field $Y$ parametrizing $S^1$
\begin{equation}
Y(z)Y(0) \sim - \ln z,~~~ 
\end{equation}
and its fermionic partner $\chi$
\begin{equation}
\chi(z)\chi(0) \sim \frac{1}{z}~~.
\end{equation}
This sector has the simplest $N = 1$ superconformal symmetry
\begin{equation}
G_{S^1} = \chi i\partial Y ~~,
\end{equation}
with the central charge 
\begin{equation}
 c = \frac{3}{2}~~.
\end{equation}
\item ${\cal N}/U(1)$ sector \\
We require that this sector has an $N =2$ superconformal symmetry
described by the currents
\begin{equation}
 T_{{\cal N}/U(1)}~,~~G^{\pm} _{{\cal N}/U(1)}~,~~J_{{\cal N}/U(1)}~~.
\end{equation}
The relation between  $N =2$ and $ N=1$  
  superconformal current is
\begin{equation}
 G _{{\cal N}/U(1)} = G^+ _{{\cal N}/U(1)} + G^- _{{\cal N}/U(1)}~~.
\end{equation}
Because of the criticality condition,
the central charge of this sector should be equal to
\begin{equation}
 c = 9 - \frac{6}{k}~,
\end{equation}
and the $U(1)_R$ current satisfies 
\begin{equation}
\dsp J_{{\cal N}/U(1)}(z) J_{{\cal N}/U(1)}(0) 
\sim \frac{3-\frac{2}{k}}{z^2}~~.
\end{equation}
\end{itemize} 
%%%%%%%%%%%%%%%%%%%%%%%%%%%%%%%%%%%%%%%%%%%%%%%%%%%%%%%%%%%%%%%%%%%%%%%%%%%

We can realize  the $N = 2 $ superconformal 
symmetry on the world-sheet in this system.
We choose the $U(1)_R$ current as
\begin{equation}
 J_R = J_{R1} + J_{R2} + J_{{\cal N}/U(1)}~,
\end{equation}
where 
\begin{eqnarray}
 J_{R1} &=& \frac{1}{2}\psi^+ \psi^- + \frac{2}{k}J^3 \label{j11}\\
%         =  \frac{k+2}{2k} \psi^+ \psi^- + \frac{2}{k} j^3 \label{j11}\\
 J_{R2} &=& \chi \psi^3~~\label{j21}.
\end{eqnarray}
According to the charge of this current
the $ N=1$ superconformal current splits into two terms
\begin{eqnarray}
 G &=& G_{SL(2,{\bf R})} + G_{S^1} + G^+_{{\cal N}/U(1)} + G^-_{{\cal N}/U(1)}
  \nonumber\\
 &\equiv& G^+ + G^-~,
\end{eqnarray}
where
\begin{eqnarray}
G^{\pm} &=& G^{\pm}_1 + G^{\pm}_2 + G^{\pm}_{{\cal N}/U(1)}\\
G^{\pm}_1 &=& \frac{1}{\sqrt{2k}}\psi^{\pm}j^{\mp} \label{g11}\\
G^{\pm}_2 &=&  \frac{1}{\sqrt{2}} (\chi \mp \psi^3)\,
  \left(\frac{1}{\sqrt{2}} i \partial Y \pm \frac{1}{\sqrt{k}}J^3\right)~~.
\label{g21}
\end{eqnarray} 
The energy-momentum tensor is also decomposed as follows;
\begin{eqnarray}
T&=& T_1+T_2+T_{{\cal N}/U(1)} \\
T_1&=& \frac{1}{k}(j^Aj_A+J^3J^3) 
-\frac{1}{4}(\psi^+\partial \psi^- - \partial \psi^+ \psi^-) \label{t11} \\
T_2 &=& -\frac{1}{k}J^3J^3 -\frac{1}{2}(\partial Y)^2
-\frac{1}{2}\chi\partial\chi +\frac{1}{2}\psi^3 \partial \psi^3. \label{t21}
\end{eqnarray} 
It may be worthwhile to mention that 
the superconformal generators $\{T_i, \,G^{\pm}_i,\, J_{Ri}\}$
(anti-) commute among the different sectors. Furthermore 
$\{T_1,\, G^{\pm}_1, \, J_{R1}\}$ has the same expression as 
that of the Kazama-Suzuki coset model for $SL(2;\br)/U(1)$ \cite{KazS}.

The BRST charge $Q_{BRST}$ is defined in the standard manner
\begin{equation}
Q_{BRST} = \oint \left[c\left(T-\frac{1}{2}(\partial \phi)^2
-\partial^2 \phi -\eta \partial \xi + \partial c b\right) 
+\eta e^{\phi} G -b\eta\partial \eta e^{2\phi}
\right],
\label{brst}
\end{equation}
where $\phi$, $\eta$, $\xi$ are the familiar bosonized superghosts \cite{FMS}.

%%%%%%%%%%%%%%%%%%%%%%%%%%%%%%%%%%%%%%%%%%%%%%%%%%%%%%%%%%%%%%%%%%%%%%%%%
Now let us try to reformulate this superstring model
as the discrete light-cone Liouville theory as in the case of bosonic string.
Our goal is the $N=2$ Liouville theory \cite{KutS1}
with the light-like compactification;
$\br_+\times S^1_- \times \br_{\rho} \times S^1 \times {\cal N}/U(1) $.
To this aim we need to perform further field redefinitions.

As a preliminary we bosonize the fermions  $\psi^{\pm}$ 
\begin{equation}
 \psi^{\pm} = \sqrt{2} e^{\pm iH_1}~,
\end{equation} 
where $ H_1(z) H_1(0) \sim - \ln z $, and the radius of compact 
boson $H_1$ should be 1.  
Let $Y_0,~Y_1$ be as given in \eqn{YYrho}, and define  
\begin{eqnarray}
 X^0 &:=&  \sqrt{\frac{k+2}{k}} Y^0 +  \sqrt{\frac{2}{k}} H_1 \nonumber\\
 X^1 &:=& - \frac{2}{\sqrt{k(k+2)}} Y^0 + \sqrt{\frac{k}{k+2}} Y^1 
          - \sqrt{\frac{2}{k}}  H_1   \label{XXH} \\
 H'_1 &:=&  \sqrt{\frac{2}{k+2}} (Y^0 + Y^1) + H_1~. \nonumber
\end{eqnarray}
Since this is again an $SO(2,1)$ rotation,
we have the  OPEs
\begin{equation}
  X^0(z)X^0(0) \sim \ln z ~,~~ X^1(z) X^1(0) \sim - \ln z ~,~~ 
 H'_1 (z) H'_1 (0) \sim - \ln z~,
\end{equation}
and all the non-diagonal OPEs vanish.
We also rewrite
\begin{equation}
 X^2 := Y~,~~\Psi^2 := \chi~,~~ \Psi^0 := \psi^3~,
 ~~~\Psi^{\pm} (\equiv -\frac{1}{\sqrt{2}}(\Psi^1\pm i\Psi^{\rho})) 
:= e^{\pm i H'_1}~~.
\end{equation}  
After all, we have changed the system of 
\begin{equation}
\{\varphi, \beta, \gamma, Y,  \psi^{\pm}, \psi^3, \chi \}
\end{equation}
into the system of new free fields
\begin{equation}
\{\rho, X^0, X^1, X^2, \Psi^{\pm}, \Psi^0, \Psi^2\}~~.
\end{equation}

In these new variables the energy-momentum tensors \eqn{t11}, \eqn{t21}
are rewritten as
\begin{eqnarray}
T_1 &=&  -\frac{1}{2}(\partial X^1)^2
-\frac{1}{2}(\partial \rho)^2 -\frac{1}{\sqrt{2k}} \partial^2 \rho
-\frac{1}{2}(\Psi^+ \partial \Psi^- - \partial \Psi^+ \Psi^-) \label{t12}\\
T_2 &=& \frac{1}{2}(\partial X^0)^2 -\frac{1}{2}(\partial X^2)^2
+\frac{1}{2}\Psi^0\partial \Psi^0 -\frac{1}{2}\Psi^2 \partial \Psi^2
\label{t22}.   
\end{eqnarray}

The $U(1)_R$ currents \eqn{j11}, \eqn{j21} become
\begin{eqnarray}
 J_{R1} &=& \Psi^+ \Psi^- - Q i \partial X^1 \label{j12}\\
 J_{R2} &=& -\Psi^0 \Psi^2~~ \label{j22} .
\end{eqnarray}
$Q$ is the background charge of Liouville mode  $\rho$ and in this case 
$\dsp Q = \sqrt{\frac{2}{k}}$.

The $N=2$ superconformal currents \eqn{g11}, \eqn{g21} now take the 
following forms
\begin{eqnarray}
G^{\pm}_1 &=& - \frac{1}{\sqrt{2}} \Psi^{\pm} 
 (i \partial X^1 \pm \partial \rho) \mp \frac{Q}{\sqrt{2}}\partial \Psi^{\pm}
\label{g12}\\
G^{\pm}_2 &=& - \frac{1}{\sqrt{2}}(\Psi^0 \mp \Psi^2) \times 
             \frac{1}{\sqrt{2}} i \partial (X^0 \pm X^2)~~.
\label{g22}
\end{eqnarray} 
It is also convenient to rewrite the total super current 
\begin{equation}
G=-\Psi^0i\partial X^0+\Psi^1i\partial X^1+\Psi^2i\partial X^2
+\Psi^{\rho}i\partial \rho + Qi\partial \Psi^{\rho}.
\end{equation}

Notice that $\{T_1,\, G^{\pm}_1,\, J_{R1}\}$ \eqn{t11}, \eqn{g11},
\eqn{j11} have been now transformed into the expressions of  superconformal
algebra in the $N=2$ Liouville theory \cite{KutS1} as we mentioned before.
The essential part of this field redefinition 
is the identification between $SL(2;\br)/U(1)$ Kazama-Suzuki model and 
the $N=2$ Liouville theory (see the appendix B of \cite{ES}, 
and also refer \cite{AFK})  and 
it was claimed in \cite{GK} that these two theories are 
related by a T-duality. 

As in the bosonic case, we introduce the light-cone coordinates
\begin{equation}
X^{\pm}=\frac{1}{\sqrt{2}}(X^0 \pm X^1),
\end{equation}
and assume the compactifications 
\begin{equation}
 X^- \sim X^- +  \frac{4 \pi}{\sqrt{k}} ,~~~ H'_1 \sim H_1' +2 \pi.
\label{llc2}
\end{equation}
These are indeed consistent with the previous compactifications
$\dsp Y^-\sim Y^-+
\frac{4 \pi}{\sqrt{k+2}}$, $H_1 \sim H_1+2 \pi$, because 
we can obtain from \eqn{XXH} 
\begin{equation}
\left\{
\begin{array}{l}
 \dsp X^- = \frac{2}{\sqrt{k(k+2)}}Y^+ + 
 \sqrt{\frac{k+2}{k}}Y^-+ \frac{2}{\sqrt{k}}H_1    \\
 \dsp H_1' = \frac{2}{\sqrt{k+2}}Y^+ + H_1  ~.
\end{array}
\right.
\end{equation}

We likewise introduce the tachyon vertices 
compatible with this light-like compactification
\begin{equation}
 V_{j,m,p} = e^{ \left( \frac{\sqrt{k}}{2} p 
 - \frac{2m}{\sqrt{k}} \right) i X^+ 
 + \frac{\sqrt{k}}{2}p i X^- - \sqrt{\frac{2}{k}}j \rho}~, 
\end{equation} 
and the corresponding Fock vacua satisfying
\begin{eqnarray}
J^3_0\ket{j,m,p}&=&(m-\frac{k}{2}p)\ket{j,m,p}, \\
J^{\pm}_{\mp p}\ket{j,m,p}&=&(m\pm j)\ket{j,m\pm 1, p}, \\
J^{\pm}_{\mp p+n}\ket{j,m,p}&=&0 , ~~~(\forall n \geq 1), \\
L_0\ket{j,m,p} &=& \left(-\frac{1}{k}j(j-1)+mp-\frac{k}{4}p^2\right)
          \ket{j,m,p} .
\end{eqnarray}

The total $SL(2;\br)$ currents are also rewritten in the new
coordinates;
\begin{eqnarray}
J^3&=& \sqrt{\frac{k}{2}}i\partial X^0 \\
J^{\pm}&=& \left(-\sqrt{\frac{k}{2}}i\partial X^1 \pm \sqrt{\frac{k}{2}}
 \partial \rho -\Psi^+\Psi^- \pm \sqrt{2} \Psi^0\Psi^{\pm}
\right)\, e^{\mp \frac{2}{\sqrt{k}}iX^+} .
\end{eqnarray}

%%%%%%%%%%%%%%%%%%%%%%%%%%%%%%%%%%%%%%%%%%%%%%%%%%%%%%%%%%%%%%%%%%%%%%
To close this section, we present the space-time superconformal
algebra in our new variables, which again has the forms reminiscent of
the DDF operators.  

First, the space-time Virasoro algebra is given (in the $(-1)$-picture) by
\begin{equation}
\left\{
\begin{array}{l}
\dsp  \cL_0 = -  \sqrt{\frac{k}{2}} \oint e^{-\phi} \Psi^0 \\ 
\dsp  {\cal L}_n = \sqrt{\frac{k}{2}}   
\oint e^{-\phi} e^{-n \frac{2}{\sqrt{k}} iX^+} 
(\Psi^1 +n i \Psi^{\rho})~~~ (n\neq 0).
\end{array}
\right.
\label{stV2}
\end{equation}
We again mention that $\cL_n ~(n\neq 0)$ is the unique solution 
of the BRST constraint among the operators of the form 
$\dsp \sim  \oint e^{-\phi} 
 e^{-n \frac{2}{\sqrt{k}} iX^+} (A\Psi^{\rho} + B\Psi^0 +C\Psi^1)$, 
$A\neq 0$.

The space-time $U(1)_R$ current is given by
\begin{equation}
  {\cal J}_n =
  \sqrt{2k}\oint  e^{-\phi} e^{-n \frac{2}{\sqrt{k}} iX^+} \Psi^2~~.
\label{stJ}
\end{equation}
%%%%%%%%%%%%%%%%%%%%%%%%%%%%%%%%%%%%%%%%%%%%%%%%%%%%%%%%%%%%%%%%%%%

To construct the space-time super currents 
we must introduce the spin fields. 
According to \cite{GRBL},
we bosonize the ``deformed  $U(1)_R$ current''  on the world-sheet;
\begin{eqnarray}
 J'_R &: =& J_R -Qi\partial (X^0+X^2)  
%\Psi^+ \Psi^- - \Psi^0\Psi^2 + (J_{{\cal N}/U(1)} -Qi\partial X^2)
          \nonumber\\
    &\equiv & i \partial H'_1 - i \partial H_2 - i \sqrt{3} \partial H_3
             - Q i\partial (X^0+X^1) ~,
\end{eqnarray}
where we set
\begin{equation}
\begin{array}{l}
 i\partial H'_1 = \Psi^+\Psi^- ~~~(\mbox{as defined before})\\
 i \partial H_2=\Psi^0\Psi^2 \\
 - i \sqrt{3} \partial H_3 = J_{{\cal N}/U(1)} -Qi\partial X^2 ~~.
\end{array}
\end{equation}
(The combined current $J_{{\cal N}/U(1)} -Qi\partial X^2$ actually has
the Schwinger term $\dsp \sim \frac{3}{z^2}$.)
The practical reason why we do so is as follows:
$J_R$ has  non-trivial OPEs with the vertex operators such as 
$ e^{-n \frac{2}{\sqrt{k}} iX^+} $, and thus the DDF like operators 
including  the  spin fields (see \eqn{f vertex}) 
made up of $J_R$ do not nicely  behave 
under the BRST transformation. In contrast, we can 
rather simply  solve the BRST condition for the vertices 
associated to the current $J'_R$,
since it has no singular OPE with $ e^{-n \frac{2}{\sqrt{k}} iX^+} $.

Now the spin fields should take the form 
(up to cocycle factors)
\begin{equation}
   \, e^{\frac{i}{2}(\epsilon_1 H'_1 
+ \epsilon_2 H_2 + \sqrt{3} \epsilon_3 H_3 )}
 e^{\epsilon_4\frac{i}{2}Q(X^0+X^1)}~~~ (\ep_i=\pm 1)~.
\end{equation}
The GSO projection leaves only a half of them satisfying
\begin{equation}
 \prod_{i=1}^3 \epsilon_i = -1~~,
\end{equation}
and we use the notation 
\begin{equation}
S^{\epsilon_3\epsilon_1} = e^{\frac{i}{2}(\epsilon_1 H'_1 
+ \epsilon_2 H_2 + \sqrt{3} \epsilon_3 H_3 )}
\end{equation}
to express the spin fields allowed by the GSO condition.
We can  explicitly verify that the fermion vertices of the type 
$\dsp \oint  e^{-\frac{\phi}{2}} e^{\frac{i}{2}\ep_4 Q(X^0+X^1)}
S^{\epsilon_3\epsilon_1}$ 
are BRST invariant, if and only if $\ep_4=-\ep_1$, namely the vertices
of the type 
$\dsp \oint  e^{-\frac{\phi}{2}} e^{-\frac{2}{\sqrt{k}}\frac{\ep_1}{2}iX^+}
S^{\epsilon_3\epsilon_1}$.
They define the space-time $N=4$ SUSY
%         \footnote
%         {Strictly speaking, in our framework of the $N=2$ Liouville
%           theory we have the SUSY only along the boundary
%          ($X^0$, $X^1$ directions) \cite{KutS1}.}
(the global part of the space-time $N=2$ superconformal symmetry 
in NS sector).
We can work out  more  general fermion vertices of the type 
\begin{equation}
 \sim \oint e^{-\frac{\phi}{2}} e^{-r\frac{2}{\sqrt{k}}iX^+} 
  \sum_{\epsilon_3,\epsilon_1} 
 A_{\epsilon_3,\epsilon_1}S^{\epsilon_3\epsilon_1} , ~~~(r\in \frac{1}{2}+\bz)
 ~.
\label{f vertex}
\end{equation} 
%which are manifestly compatible with the light-like compactification.
The BRST condition uniquely (up to BRST exact terms and an  overall
normalization) determines the coefficients $A_{\epsilon_3,\epsilon_1}$
and we finally obtain the physical vertices
\begin{equation}
  {\cal G}_r^{\pm} = \dsp  k^{1/4}
  \oint e^{-\frac{\phi}{2}} e^{-r \frac{2}{\sqrt{k}} iX^+} \left[
    (r+\frac{1}{2})S^{\pm +} -  (r-\frac{1}{2})S^{\pm -}  \right] ~,
~~~ (r \in \frac{1}{2}+\bz)~.
\label{stG}
\end{equation}
They generate the $N=2$ superconformal algebra (in NS sector)
together with $\cL_n$, $\cJ_n$, and  the central charge 
is equal to $6kp$ on the Fock space over the vacuum $\ket{*,*,p}$. 
In fact, it is a straightforward calculation to check
that these space-time superconformal generators 
\eqn{stV2}, \eqn{stJ}, \eqn{stG} coincide with  
the ones  constructed in \cite{GRBL} up to BRST exact terms.

%%%%%%%%%%%%%%%%%%%%%%%%%%%%%%%%%%%%%%%%%%%%%%%%%%%%%%%%%%%

~

\section{Analyses on Spectra of Physical States}

\hspace*{4.5mm}

In this section we shall investigate the spectrum of physical states
in our reformulation of $AdS_3$ string theory. 
%We assume $k\geq 1$ through this section. 
The unitarity of physical Hilbert space is an important problem.
We first analyze the physical spectrum as that of {\em a free field theory}, 
namely without taking the Liouville potential term into account, 
and later discuss the effect of turning on this term.

\subsection{Spectrum as Free Field Theory}

\hspace*{4.5mm}

First, we analyze the spectrum of bosonic string on $AdS_3 \times {\cal N}$. 
Let us recall that the  Fock vacuum is defined from the tachyon vertex
\begin{equation}
\ket{j,m,p}= \lim _{z \to 0} V_{j,m,p}(z)\ket{0}~~,
\end{equation}
and we denote the corresponding Fock space as $\cF_{j,m,p}$ from now on.  
We also define ``out state'' as
\begin{equation}
 \bra{j,m,p} = \lim _ {z \to \infty} \bra{0} V_{j,m,p} (z) z^{2 h_{j,m,p}}~,
\end{equation}
where
\begin{equation}
 h_{j,m,p} = - \frac{1}{k} j ( j-1) + mp - \frac{k+2}{4} p^2~~.
\end{equation}
Using the momentum conservation and taking account of 
the existence of background charge along the $\rho$-direction, 
we obtain
\begin{equation}
 \langle 1-j ,-m,-p \ket{j,m,p}  \neq 0~,
\end{equation}
and the other combinations vanish.
Notice that we must use the following Hermitian conjugation
\begin{equation}
\bra{1-j,-m,-p}=\left(\ket{j,m,p}\right)^{\dag}
\end{equation}
to discuss the unitarity.

We shall here neglect the Liouville potential term (that is, the ``screening
charge term'' 
$\dsp \int \beta\bar{\beta}e^{-\sqrt{\frac{2}{k}}\varphi} \sim 
\int \partial Y^1 \deebar Y^1 \, e^{-\sqrt{\frac{2}{k}}\rho}$
in the $\sigma$-model action \eqn{ads3 lagrangian}).
This means that the Hilbert space of physical states should be 
defined as the BRST cohomology on the Fock spaces 
of the free fields $Y^0$, $Y^1$, $\rho$ properly tensored by 
the Hilbert space of $\cN$ sector.  
Deciding the physical spectrum is 
a rather simple problem as in  the usual string theory on the flat
Minkowski space (at least as long as we only take  the primary
states in the $\cN$ sector in constructing the physical states). 
However, there is one non-trivial point: the existence of background 
charge in the $\rho$-direction.  As is well-known, 
the BRST constraint eliminates two longitudinal degrees of freedom
in the case of Minkowski background:
one is eliminated by the BRST condition itself and another becomes BRST exact. 
%Only the transverse oscillators can span the physical Hilbert space. 
From a physical point of view  this aspect is closely related 
to the fact that the first excited states (the graviton states)
become mass-less, and thus one of the polarization vectors must be
light-like. On the other hand, in our case of linear dilaton theory 
we have a mass gap originating from the background charge of $\rho$
and so the first excited states become {\em massive\/}. 
This implies that one of the two longitudinal modes 
does {\em not\/} decouple from the physical Hilbert space, since
all the polarization vectors are space-like.

To make this point clearer, let us consider a simple example.
We are here given one transverse oscillator 
$\dsp i\partial \rho = \sum_n \frac{\al^{\rho}_n}{z^{n+1}}$
and two longitudinal oscillators
$\dsp i\partial Y^{\pm} = \pm \sum_n \frac{\al^{\pm}_n}{z^{n+1}}$.
The simplest candidate of the first excited states is 
\begin{equation}
\al^{\rho}_{-1} \ket{p^+, p^-, p^{\rho}} \otimes c_1 \ket{0}_{\msc{gh}} ,
\label{ex1}
\end{equation} 
where the on-shell condition is given by 
\begin{equation}
p^+p^- +\frac{1}{2}(p^{\rho})^2 + \frac{1}{4k}=0 ~~.
%\left(Q=\sqrt{\frac{2}{k}}\right) .
\end{equation}
(The relation of the momenta with our previous convention is given by
$\dsp p^+ =-\frac{\sqrt{k+2}}{2}p$, 
$\dsp p^-=\frac{\sqrt{k+2}}{2}p - \frac{2m}{\sqrt{k+2}}$, 
$\dsp p^{\rho}=i\sqrt{\frac{2}{k}}\left(j-\frac{1}{2}\right)$.)
Now we assume $p^+p^-\neq 0$. 
The BRST transformation of this candidate \eqn{ex1} yields a
non-vanishing term due to the background charge.  
We must cancel it by mixing the longitudinal modes to 
recover the BRST invariance. After some simple calculation
we find  {\em two\/} independent solutions
\begin{equation}
\left(\al^{\rho}_{-1}+i \frac{Q}{p^{\pm}}\al^{\pm}_{-1}\right)
\ket{p^+,p^-,p^{\rho}}\otimes c_1\ket{0}_{\msc{gh}}.
\label{ex2}
\end{equation}

In the usual free string theory the general physical states
are created by making the transverse DDF operators act on suitable Fock vacuum
(``allowed states'').
The above simple observation suggests that, in our case of $AdS_3$,
we must make use of the two independent DDF operators 
that are {\em not purely transversal\/}.
Some candidates for the suitable DDF operators
are already given by Satoh \cite{Satoh}\footnote
    {In \cite{Satoh} the other candidates for the DDF operators are
    also proposed.  
   However, they  include the ghost number current explicitly in their
   expressions and are not BRST invariant. 
   Y.S. should express his great thanks to Dr. Satoh 
   for the  discussion about this point.}
\begin{equation}
B^{\pm}_n :=\oint i
\left(\partial \rho+\frac{Q}{2}\partial \ln \partial Y^{\pm}\right) 
e^{\frac{n}{p^{\pm}}iY^{\pm}},
\end{equation}
which are BRST invariant and satisfy the commutation relation of a free
boson 
\begin{equation}
\begin{array}{l}
[B_m^{\pm}, ~B_n^{\pm}]=m\delta_{m+n,0}, \\
B_m^{\pm} \ket{p^+,p^-,p^{\rho}}=0, ~~~(\forall m\geq 1)
\end{array}
\end{equation}
Moreover,  $B^{\pm}_n$ act on the Fock space 
defined by $\ket{p^+,p^-,p^{\rho}}$ as 
the operators $\sim \al^{\rho}_{n}+ \cdots$,  as is expected, and it is easy 
to check that $B_{-1}^{\pm}$ actually give rise to the first excited 
states discussed above \eqn{ex2}.
So the reader might suppose  that we can  naively use $B^+_n$, $B^-_n$ 
to construct the complete set of physical states.
But this is not the case. $B^+_n$ and $B^-_m$ are {\em not
mutually local\/}  and thus we can never use them at the same time.
The best we can do is to use  only one of them, $B^+_m$, 
which is compatible with the light-like compactification \eqn{llc}. 
We rewrite it as 
\begin{equation}
 \cA^{(p)}_n = \oint
  i\left(\partial \rho + \frac{Q}{2}\partial 
     \ln \partial Y^+\right)
    e^{-\frac{n}{p}\frac{2}{\sqrt{k+2}}i Y^+}~,~~~
 {[} \cA^{(p)}_m, \cA^{(p)}_n {]} = m \delta_{m+n,0}~~.
\label{DDFA}
\end{equation}
which are defined as local operators on $\cF_{*,*,p}$ ($p\neq 0$).

Now, the question we must solve is as follows:
What is the missing DDF operator that can compensate 
\eqn{DDFA}?  
As we already suggested, the answer is 
the space-time Virasoro operators.
Let $\cL_n$ be the space-time Virasoro operators defined 
in \eqn{stV1}. $\cL_{\frac{n}{p}}$ are well-defined as local operators
on the Fock space $\cF_{j,m,p}$ and behave as 
$\sim \al^{\rho}_{n}+ \cdots$. Therefore they are the candidates of 
the missing DDF operators. Alternatively we shall define   
\begin{equation}
 \cL^{(p)}_n 
:= p  {\cal L}_{\frac{n}{p}} - \frac{k+2}{4} (p^2-1) 
 \delta_{n,0} ~,
\end{equation}
which are shown to generate a Virasoro algebra  
with the central charge $c=6(k+2)$ (irrespective of the value $p$).
Furthermore, $\cL^{(p)}_n$ are  mutually local with $\cA^{(p)}_m$
and satisfy the commutation relation
\begin{equation}
[\cL^{(p)}_n, ~\cA^{(p)}_m ] = -m \cA^{(p)}_{m+n} +i \al n^2\delta_{n+m,0} ,
~~~\al\equiv \sqrt{\frac{k}{2}}(1-\frac{1}{k}).
\label{LA} 
\end{equation}
It is also convenient to introduce improved Virasoro operators
\begin{equation}
\tilde{\cL}_n^{(p)} := \cL_n^{(p)} 
-\frac{1}{2} \sum_m :\cA^{(p)}_{-(n+m)}\cA^{(p)}_m: 
- i\al n \cA^{(p)}_n
-\frac{1}{2}\al^2 \delta_{n,0} ~,
\label{tildeL}
\end{equation} 
which are defined so that they commute with $\{\cA^{(p)}_m\}$
and generate the Virasoro algebra with 
$\dsp c= 23-\frac{6}{k} \equiv c_{\cN}$.
This value of the  central charge is quite expected. 
One can show that the DDF operators $\tilde{\cL}^{(p)}_n$ correspond to 
the energy-momentum tensor of $\cN$ sector in the light-cone gauge.

In this way, we have found that the physical Hilbert space should be
spanned by the states having the forms
\begin{equation}
\tilde{\cL}^{(p)}_{-n_1}\tilde{\cL}^{(p)}_{-n_2}\cdots
\cA^{(p)}_{-m_1}\cA^{(p)}_{-m_2}\ket{j,m,p} \otimes \cdots ~, 
~~~(n_i\geq 1, m_i\geq 1)~.
\label{phys}
\end{equation}

In order to complete our discussion  
we must confirm the linear independence of the states of the type 
\eqn{phys}. Happily, this is very easy to prove in our case.  
The Virasoro algebra $\{\tilde{\cL}_n^{(p)}\}$ has 
the central charge greater than 1
for sufficiently large $k$, and thus the Kac determinant 
does not vanish, as is well-known in the representation theory of 
Virasoro algebra.
%\footnote
%      {Strictly speaking,  one must perform  a further improvement 
%$$
%\tilde{\cL}_n^{'(p)} := \cL_n^{(p)} 
%-\frac{1}{2} \sum_m :\cA^{(p)}_{-(n+m)}\cA^{(p)}_m: 
%- i\al n \cA^{(p)}_n
%-\frac{1}{2}\al^2 \delta_{n,0}
%$$
%to make them  commute with $\{\cA^{(p)}_m\}$. These Virasoro algebra
%has central charge $\dsp c= 23-\frac{6}{k}$.}. 

%%%%%%%%%%%%%%%%%%%%%%%%%%%%%%%%%%%%%%%%%%%%%%%%%%%%%%%%%%%%%%%%%%%%%%%

We can now present the complete list of physical states. 
This spectrum is specified by the momenta of the Fock space 
$\cF_{j,m,p}\otimes \bar{\cF}_{j,\bar{m},p}$ previously defined. 
The light-like compactification \eqn{llc} implies that 
$p\in \bz$, and also $\tilde{\cL}^{(p)}_0
-\overline{\tilde{\cL}^{(p)}_0}\equiv
p(\cL_0-\bar{\cL}_0) \in p\bz$ (the ``level matching condition'').
Since $\cA^{(p)\dag}_0= \cA^{(p)}_0$ holds and we have
\begin{equation}
\cA^{(p)}_0 \ket{j,m,p}= 
i\sqrt{\frac{2}{k}}\left(j-\frac{1}{2}\right)\ket{j,m,p} ,
\label{eigen-A0}
\end{equation}
the value of $j$ allowed by the unitarity is $\dsp j=\frac{1}{2}+is$
($s\in \br$), at least when $p\neq 0$.
It corresponds to the principal continuous series of 
unitary representation of $SL(2;\br)$. Also in the case of $p=0$ 
we can show that only the principal series is permitted from 
the unitarity, as we will observe below.

To avoid unnecessary  complexity
we shall only  take the primary states in the $\cN$ sector
in constructing the physical states,
and assume the conformal weights $h_{\cN}$ of these primary states 
are non-negative. It is not difficult to construct more general 
physical states including the descendant states in the $\cN$ sector,
if we are concretely given the unitary CFT  model describing  this sector.    

We  discuss the $p=0$ and $p\neq 0$ cases separately.

\begin{enumerate}
\item  $p=0$ \\
In this case, the DDF operators of the types 
$\cA^{(p)}_n$, $\cL^{(p)}_n$ are ill-defined. But we must require
the space-time Virasoro operators $\{\cL_n\}$ (with central charge 0) 
should define an unitary representation,
since they are well-defined as local operators even in this sector. 
%although 
%it is neither a highest weight nor a lowest weight representation.   
$\cL_n$ simply maps a Fock vacuum 
to another  Fock vacuum and  so
the representation space is given by 
$\dsp \oplus_{r\in \bsz}\,\bc\ket{j,m+r,0}$
with arbitrary fixed values of $m\in \br$, $j$.
We can obtain 
\begin{equation}
\begin{array}{l}
\dsp \bra{1-j, -m-r, -p}\cL_{-n}\cL_n\ket{j,m+r,p}\\
\dsp   \hspace{1.5cm}= \left\{\left(m+r+\frac{n}{2}\right)^2
      -n^2\left(j-\frac{1}{2}\right)^2\right\}
\,\langle 1-j,-m-r-n,0|j,m+r+n,0\rangle .
\end{array}
\label{p0-eq}
\end{equation}
Since the conformal weight (in the sense of world-sheet CFT) must be 
real at least, $j$ should take the values  
$j\in \br$ or $\dsp j=\frac{1}{2}+is$ $(s\in \br)$.
If $\dsp j-\frac{1}{2} \in i\br$ (principal series), 
the coefficient of R.H.S in \eqn{p0-eq} 
is always  positive and we have an unitary representation
of $\{\cL_n\}$. 
On the other hand, if $j\in \br$,  we can always find $r\in \bz$ 
for which this coefficient becomes negative as long as we choose
$n$ to be sufficiently large.
This means that the cases of $j\in \br$ cannot be 
unitary representations of $\{\cL_n\}$, and hence we must 
rule out these sectors.
 
The general physical states with $p=0$ are written as 
\begin{eqnarray}
 \lefteqn{\ket{\frac{1}{2}+is,m,0} \otimes \ket{h_{\cal N}} 
 \otimes c_1 \ket{0}_{\msc{gh}}}\nonumber\\
 &\otimes&  \ket{\frac{1}{2}+is,\bar{m},0} \otimes \ket{h_{\cal N}} 
 \otimes \bar{c}_1 \ket{0}_{\msc{gh}}\\
 &~&m,\bar{m} \in \br~,~m-\bar{m}\in  \bz \nonumber~,
\end{eqnarray}
where $\ket{h_{\cal N}}$ is the primary state with conformal weight 
  $h_{\cal N}$ in the ${\cal N}$ sector 
  and $\ket{0}_{\msc{gh}}$ is the  vacuum of the ghost system. 
The on-shell condition is given by
\begin{equation}
 \frac{1}{k}\left(s^2 + \frac{1}{4}\right) + h_{\cal N}= 1~~.
\label{on-shell1}
\end{equation}
If $k>1/4$, we can always solve the on-shell condition \eqn{on-shell1}
for $h_{\cal N}=0$. These physical states are tachyons  whose 
mass-squared are lower than the Breitenlohner-Freedman bound \cite{BrF}.
Such an instability in bosonic string theory is not surprising,
and we later observe that the GSO projection successfully 
eliminates these tachyonic states in the superstring case.

There is one comment: If we took  account of the unitarity of 
the representation {\em only of  $SL(2;\br)$} (that is, $\{\cL_n\}$, 
$n=0,\, \pm 1$), many representations would survive in the sector
$j\in \br$: the discrete series $\cD^{\pm}_j$ and 
the exceptional series $\cE_{j,\al}$, as is
well-known and many readers might expect.
It is crucial in the above argument  to take the {\em full\/}
Virasoro algebra $\{\cL_n\}$ in place of the $SL(2;\br)$ subalgebra. 
%%%

\item $p \neq 0$\\
As we already discussed, in this sector $j$ must be equal to 
$\dsp j=\frac{1}{2}+is$ $(s\in \br)$, and 
the physical Hilbert space is 
generated by the actions of the DDF operators
$\{\cA^{(p)}_{-n}\}$, $\{\tilde{\cL}^{(p)}_{-n}\}$ $(n\in\bz_{\geq 1})$
on the on-shell Fock vacua. 
We must discuss the positivity of the norm of such physical states.
Obviously $\cA^{(p)}_{-n}$ create only positive norm states,
and do not lead to any constraint.  However, the Virasoro generators
$\{\tilde{\cL}^{(p)}_n\}$ give rise to a non-trivial constraint for
unitarity. Since this Virasoro algebra has the central charge $c>1$,
the condition for the unitarity means that the 
$\tilde{\cL}^{(p)}_0$-eigenvalue of the Fock vacuum
$\dsp \ket{\frac{1}{2}+is,m,p}$
is non-negative.
(We again assume $k$ is sufficiently large.)
It is easy to show that this unitarity condition 
is equivalent to a simple inequality  $h_{\cN}\geq 0$ thanks to
the on-shell condition
\begin{equation}
\frac{1}{k} \left( s^2 + \frac{1}{4}\right) + mp - \frac{k+2}{4}p^2 +
	 h_{\cal N} = 1~~. 
\label{on-shell-p}
\end{equation}
This equivalence is not surprising, since $\tilde{\cL}^{(p)}_0$ corresponds 
to the stress tensor for $\cN$ sector in the light-cone gauge.  
In this way we conclude that the no-ghost theorem
for this sector  is trivially satisfied as long as 
the internal CFT $\cN$ is unitary. 
This result is consistent with those of \cite{Bars,Satoh}, although
we here take a different convention of free field representation:
Our convention diagonalizes the time-like current $j^3$ (corresponding
to the energy operator). On the other hand, those given 
in \cite{Bars,Satoh,BDM} diagonalize one of the space-like currents. 
We remark that the light-cone momentum $p$ plays a role similar to 
that of the extra zero-mode momentum emphasized in \cite{Bars,Satoh,BDM}.

One can find that the $\cL_0$-eigenvalue (not $\tilde{\cL}_0^{(p)}$)
of the on-shell Fock vacuum,
which corresponds to the space-time energy,
is bounded below
\begin{equation}
 \cL_0 \geq \frac{h_{\cN}}{p} + \frac{(k-1)^2}{4kp}+
 \frac{k+2}{4}\left(p-\frac{1}{p}\right) \sim 
\frac{h_{\cN}-1}{p}+\frac{k+2}{4}p,
\end{equation}
(for a sufficiently large value $k$). 
This means that this sector corresponds to the long string states 
in the sense of \cite{MO}
and belongs to the continuous spectrum 
above the threshold energy $\dsp \sim \frac{k}{4}p $ discussed in 
\cite{MMS,SW}.

In summary, the general  physical states are written as 
\begin{eqnarray}
 \lefteqn{\tilde{{\cL}}^{(p)}_{-n_1}\tilde{{\cL}}^{(p)}_{-n_2}\cdots
 \cA^{(p)}_{-m_1}\cA^{(p)}_{-m_2}\cdots
 \ket{\frac{1}{2}+is,m,p} \otimes \ket{h_{\cal N}} 
 \otimes c_1 \ket{0}_{\msc{gh}}} \nonumber \\
 &\otimes& 
 \overline{\tilde{{\cL}}^{(p)}}_{-\bar{n}_1}
 \overline{\tilde{{\cL}}^{(p)}}_{-\bar{n}_2}\cdots
 \overline{\cA^{(p)}}_{-\bar{m}_1}\overline{\cA^{(p)}}_{-\bar{m}_2}\cdots
 \ket{\frac{1}{2}+is,\bar{m},p} \otimes \ket{\bar{h}_{\cal N}} 
 \otimes \bar{c}_1 \ket{0}_{\msc{gh}} \nonumber\\
 && n_1,n_2,\cdots \geq 1 ,~~~m_1,m_2,\cdots \geq 1 \nonumber \\
&& \bar{n}_1,\bar{n}_2,\cdots  \geq 1,~~~\bar{m}_1,\bar{m}_2,\cdots  \geq 1~~,
% ~ \sum_i n_i + \sum_j m_j  - mp 
% = \sum_i \bar{n}_i + \sum_j \bar{m}_j- \bar{m}p ~~(\mod p)\nonumber~~,
\label{phys-pr}
\end{eqnarray}
where the on-shell conditions are
 \begin{eqnarray}
 \frac{1}{k}\left(s^2 + \frac{1}{4}\right) + mp
  - \frac{k+2}{4}p^2 + h_{\cal N}&=& 1 \nonumber\\
 \frac{1}{k}\left(s^2 + \frac{1}{4}\right) + \bar{m}p
  - \frac{k+2}{4}p^2 + \bar{h}_{\cal N}&=& 1 ~~,
\end{eqnarray}
and the ``level matching condition'' is given as 
\begin{equation}
\sum_i n_i + \sum_j m_j  - mp 
= \sum_i \bar{n}_i + \sum_j \bar{m}_j- \bar{m}p ~~(\mod ~ p)~~.
\end{equation}

\end{enumerate}

%%%%%%%%%%%%%%%%%%%%%%%%%%%%%%%%%%%%%%%%%%%%%%%%%%%%%%%%%%%%%%%%%%%%%%

~

The superstring case $AdS_3 \times S^1 \times \cN/U(1)$ is similarly analyzed. 
The unitarity of the physical Hilbert space is derived  
from  the unitarity in the $N=2$ SCFT describing $\cN/U(1)$ sector.
We here only discuss how tachyonic states in the $p=0$ sector
are eliminated by the GSO projection.

The tachyon vertex operators have  slightly different expressions 
as compared with the bosonic case 
\begin{equation}
 V_{j,m,p} = e^{ \left( \frac{\sqrt{k}}{2} p 
 - \frac{2m}{\sqrt{k}} \right) i X^+ 
 + \frac{\sqrt{k}}{2}p i X^- - \sqrt{\frac{2}{k}}j \rho}~.
\end{equation} 
Together with the vertex for the $S^1$ direction
$\dsp e^{-i\sqrt{\frac{2}{k}}qX^2}$ 
we construct  the Fock vacuum $\ket{j,m,p,q}$ such that 
\begin{eqnarray}
J^3_0\ket{j,m,p,q}&=&(m-\frac{k}{2}p)\ket{j,m,p,q}, \\
J^{\pm}_{\mp p}\ket{j,m,p,q}&=&(m\pm j)\ket{j,m\pm 1, p,q}, \\
J^{\pm}_{\mp p+n}\ket{j,m,p,q}&=&0 , ~~~(\forall n \geq 1), \\
L_0\ket{j,m,p,q} &=& \left(-\frac{1}{k}j(j-1)+mp-\frac{k}{4}p^2
          +\frac{q^2}{k}  \right) \ket{j,m,p,q} , \\
J'_{R0}\ket{j,m,p,q} &=& \left(\frac{2q}{k}+p\right)\ket{j,m,p,q}.
\end{eqnarray}

In the $p=0$ sector the argument similar to the bosonic case leads 
to the following  physical states;
\begin{eqnarray}
 \lefteqn{\ket{\frac{1}{2}+is,m,0,q} \otimes \ket{h_{\cal N}, q_{\cN}} 
 \otimes c e^{- \phi}\ket{0}_{\msc{gh}}}\nonumber\\
 &\otimes&  \ket{\frac{1}{2}+is,\bar{m},0, \bar{q}} 
             \otimes \ket{\bar{h}_{\cal N},\bar{q}_{\cN}} 
 \otimes \bar{c}  e^{- \bar{\phi}}\ket{0}_{\msc{gh}}\\
 &~&m,\bar{m} \in \br~,~m-\bar{m}\in  \bz \nonumber~,
\end{eqnarray}
with the on-shell conditions
\begin{equation}
\begin{array}{l}
\dsp  \frac{1}{k}\left(s^2 + \frac{1}{4}\right) + \frac{q^2}{k}+
   h_{\cal N}= \frac{1}{2} \\
\dsp  \frac{1}{k}\left(s^2 + \frac{1}{4}\right) + \frac{\bar{q}^2}{k}+
   \bar{h}_{\cal N}= \frac{1}{2}~.
\end{array}
\label{on-shell-s}
\end{equation}
Naively we can solve the on-shell conditions as in the bosonic case
and they are tachyonic states except for $s=0$. 
However, we can show that the GSO projection eliminates 
such tachyonic states, as is expected. 
From the on-shell conditions \eqn{on-shell-s} and the assumption
$\dsp h_{\cN}\geq \frac{1}{2}|q_{\cN}| $, which is derived from the 
unitarity in the $\cN/U(1)$-sector,  
we obtain the inequality
\begin{equation}
\frac{1}{k}(\frac{1}{4}+s^2) +\frac{q^2}{k} +\frac{|q_{\cN}|}{2}
\leq \frac{1}{2} .
\label{ineq1}
\end{equation}
We should define the GSO condition
with respect to the deformed $U(1)_R$ current $J'_R$ 
and it reads as $\dsp \frac{2q}{k}+q_{\cN}=2l +1 $  
($l\in \bz$). 
First we assume $q_{\cN} \geq 0$.
If $l\geq 0$, substituting $\dsp q_{\cN}=2l+1-\frac{2q}{k}$
into the above inequality
\eqn{ineq1}, we obtain
\begin{equation}
s^2+\left(q-\frac{1}{2}\right)^2 +2lk \leq 0,
\end{equation}
which leads to $n=0$, $\dsp q=\frac{1}{2}$, $s=0$.
In the case of  $l<0$,  
$\dsp q\leq \frac{k}{2}(2l+1)$ must hold.
We thus  obtain
\begin{eqnarray}
 \frac{1}{2}&\geq& \dsp \frac{1}{k}\left(\frac{1}{4}+s^2\right)
 + \frac{q^2}{k}+\frac{|q_{\cN}|}{2}
 \, \geq \,  \frac{s^2}{k} + \frac{1}{2},
\end{eqnarray}
which leads to $s=0$, again.
Therefore the tachyonic states whose mass-squared are lower than 
the BF bound are successfully eliminated by the GSO projection. 
We can repeat the same analysis when  $q_{\cN} < 0$.

~

%%%%%%%%%%%%%%%%%%%%%%%%%%%%%%%%%%%%%%%%%%%%%%%%%%%%%%%%%%%%%%%%%%%%%
\subsection{Physical Spectrum under the Existence of Liouville
Potential}

\hspace*{4.5mm}

In the previous argument only the principal series was  allowed.
In the physical sense it was a quite natural result, because 
we regarded the system as a {\em free system\/}   and 
thus all the momenta should be real.
% from the requirement of unitarity. 

Now, let us try to turn on the Liouville potential term
(or the screening charge term in the world-sheet action 
\eqn{ads3 lagrangian}). In this case we can expect 
some physical states with an imaginary
momentum along the $\rho$-direction describing the bound 
states (``bound string states'' in the terminology of \cite{MO}).

Unfortunately,  a rigorous treatment of the quantum Liouville theory as an 
interacting theory is quite non-trivial. 
Instead we shall here take the operator contents as free fields and 
treat the Liouville potential as a small perturbation.

Recalling the $\sigma$-model action \eqn{ads3 lagrangian}, 
this perturbation term may be identified with the 
operator $\sim S\bar{S}$, where    
\begin{equation}
S = \oint \beta e^{-\sqrt{\frac{2}{k}}\varphi}
\equiv -\sqrt{\frac{k+2}{2}}\oint i\partial Y^1 e^{-\sqrt{\frac{2}{k}}\rho}
\label{sc1}
\end{equation}
is no other than the familiar screening charge 
which commutes with all modes of $\widehat{SL}(2;\br)$ currents.
As for the spectrum generating operators, we may as well expect that
at least $\cL_{\pm 1} (\sim \cL^{(p)}_{\pm p}), \,  \cL_0$ 
remain the  good DDF operators, 
since they commute with the screening charge \eqn{sc1}.

On the other hand,  because such an interaction breaks the translational 
invariance along the $\rho$-direction, the $\rho$-momenta 
$\sim \dsp i\left(j-\frac{1}{2}\right)$ loses its  meaning. 
However the second Casimir 
%of $\{\cL_{\pm 1},\,\cL_0\}$;  
$\sim j(j-1)$ remains  well-defined as 
a conserved quantity characterizing the physical
states even under the interacting theory. This is nothing but 
the standard argument of ``screening out'' of  the extra 
zero-mode momentum  in the free field representation of $CFT_2$
\cite{DF,BF,Wakimoto}. We may expect the   
bound string states possessing the imaginary $\rho$-momenta
as long as their  second Casimirs  take real values.
In this way  we can no longer regard $\cA^{(p)}_0$ 
as a good DDF operator.  
Moreover, we must also rule out the non-zero modes $\cA^{(p)}_n$ $(n\neq 0)$,
because we have the following  commutation relations 
$\cA^{(p)}_0 \sim [ (\cL_{-1})^n,\, \cA^{(p)}_{n} ] +\mbox{const.~}$
for $n>0$, and 
$\cA^{(p)}_0 \sim [ (\cL_{1})^{-n},\, \cA^{(p)}_{n} ] +\mbox{const.~}$
for $n<0$.

It is a subtle problem whether 
the other modes of Virasoro operators $\cL^{(p)}_n$ ($n\neq 0,\, \pm p$) 
remain the members of the spectrum generating algebra, 
since they also do not commute with the screening charge
\eqn{sc1}.  However, it may be plausible to admit these operators  
from the point of view of the $AdS_3/CFT_2$ correspondence 
or the arguments of Brown-Henneaux
\cite{BH}. The fact that only the $SL(2;\br)$ generators 
$\cL_0,\, \cL_{\pm 1}$ commute with the screening charge and 
the other modes do not is supposed to reflect the following fact:
In the argument of \cite{BH} the true isometry generates only the $SL(2;\br)$ 
and the other modes merely correspond to the asymptotic isometries,
which can be regarded as symmetries only near the boundary.       
We shall now  propose  that 
{\em the DDF operators suitable for the interacting theory including 
\eqn{sc1} are $\{\cL^{(p)}_n \}$ rather than those for the free system
$\{\tilde{\cL}^{(p)}_n , \, \cA^{(p)}_n\}$.}
This claim is consistent with  the analyses  based on the light-cone gauge
for the long string configuration \cite{YZ,SW}. Hence  our assertion 
is likely  to be consistent at least with the spectrum of 
the long string located near the boundary. (One should keep it in mind that 
our assumption of small Liouville perturbation is valid only 
for such a configuration of world-sheet.)

%%%%%%%%%%%%%%%%%%%%%%%%%%%%%%%%%%%%%%%%%%%%%%%%%%%%%%%%%%%%%%%%%%%%%%%%%%

Based on this assumption we now present the complete physical 
spectrum as the interacting theory.
For the principal series $\dsp j=\frac{1}{2}+is$, 
the results are similar to those of  free fields.
The case of $p=0$ is the same as before, and when $p\neq 0$, 
only we have to do is to replace the DDF operators 
$\{\cA^{(p)}_n,\, \tilde{\cL}^{(p)}_m\}$ in the expression 
of \eqn{phys-pr} by $\{\cL_n^{(p)}\}$. 
They likewise belong to a continuous spectrum  above the threshold
energy $\dsp \sim \frac{kp}{4}$.

A crucial difference is the existence of the physical states 
with $j\in \br$ as we already suggested.
To discuss the unitarity of this sector
we again assume $p > 0$, and the cases of $p<0$ can be analyzed in the same
way.  The unitarity condition means that  the $\cL^{(p)}_0$-eigenvalue
of the Fock vacuum should be non-negative. Thanks to the on-shell
condition
\begin{equation}
 -\frac{1}{k}j(j-1) +mp - \frac{k+2}{4}p^2 + h _{\cal N} = 1~,
\label{on-shell2}
\end{equation}
we can immediately obtain the following inequality for  
the unitarity
\begin{equation}
 \frac{1}{k} \left(j-\frac{1}{2}\right)^2 \leq 
 h_{\cN} + \frac{(k-1)^2}{4k} ~~.
\label{unitarity-ineq2}
\end{equation}
We must also restrict the range of $j$ as
$j>1/2$ because of the normalizability of wave function (see \cite{KutS2}).  
Especially, in the case of $h_{\cN}=0$ we obtain the unitarity condition
\begin{equation}
\frac{1}{2}<j\leq \frac{k}{2}. 
\label{j-restriction}
\end{equation}
These physical states do not propagate along the radial direction
$\rho$, and are supposed to correspond to the bound string states
in the argument of \cite{MO}. In fact, we can evaluate the space-time 
energy for this sector
\begin{equation}
 \frac{k+2}{4}\left(p-\frac{1}{p}\right) 
 \lsim \cL_0 \lsim \frac{h_{\cN}-1}{p} + \frac{k+2}{4}p ,  
\label{energy-ineq}
\end{equation}
which is consistent with the result given in \cite{MO}.

The physical states are summarized as follows;
\begin{eqnarray}
 \lefteqn{\cL^{(p)}_{-n_1}\cL^{(p)}_{-n_2}\cdots
 \ket{j,m,p} \otimes \ket{h_{\cal N}} 
 \otimes c_1 \ket{0}_{\msc{gh}}} \nonumber \\
 &\otimes& 
 \overline{\cL^{(p)}_{-\bar{n}_1}} \,
 \overline{\cL^{(p)}_{-\bar{n}_2}}\cdots
 \ket{j,\bar{m},p} \otimes \ket{\bar{h}_{\cal N}} 
 \otimes \bar{c}_1 \ket{0}_{\msc{gh}} \nonumber\\
&& n_1,n_2,\cdots \geq 1 ,~~~ 
    \bar{n}_1,\bar{n}_2,\cdots  \geq 1,~~~
% ~ \sum_i n_i - mp 
% = \sum_i \bar{n}_i - \bar{m}p ~~(\mod p)\nonumber~~,
\end{eqnarray}
where the on-shell conditions are
 \begin{eqnarray}
- \frac{1}{k}j(j-1) + mp
  - \frac{k+2}{4}p^2 + h_{\cal N}&=& 1 \nonumber\\
- \frac{1}{k}j(j-1) + \bar{m}p
  - \frac{k+2}{4}p^2 + \bar{h}_{\cal N}&=& 1 ~~.
\end{eqnarray}
and the ``level matching condition'' is given as 
\begin{equation}
\sum_i n_i - mp 
= \sum_i \bar{n}_i - \bar{m}p ~~(\mod ~ p).
\end{equation}

%%%%%%%%%%%%%%%%%%%%%%%%%%%%%%%%%%%%%%%%%%%%%%%%%%%%%%%%%%%%%%%%%%%%%%%%%
We here remark that the positive (negative) 
energy ($\cL_0\geq 0$) physical states 
should have $p>0$ ($p<0$). This fact will be important 
in the discussions in the next section about
the interpretation of the long string theory.

%%%%%%%%%%%%%%%%%%%%%%%%%%%%%%%%%%%%%%%%%%%%%%%%%%%%%%%%%%%%%%%%%%%%%%%%%
To close this section we compare the above spectrum with the result of  
\cite{MO}. For this purpose we must clarify which representations of 
$\widehat{SL}(2;\br)$ we should choose. 

Let us assume $p<0$. Going back to the Wakimoto free fields 
$\varphi$,  $\beta$, $\gamma$ we introduce the ``Wakimoto module''
$\cW_{j,m,p}$ which  is defined as the Fock space generated 
by $\al^{\varphi}_{-n-1}$, $\beta_{p-n}$, $\gamma_{-p-n}$ $(n\geq 0)$
out of the vacuum $ \ket{j,m,p}$ for $m \neq j$, and
by $\al^{\varphi}_{-n-1}$, $\beta_{p-n-1}$, $\gamma_{-p-n}$ $(n\geq 0)$ 
for $m=j$ (corresponding to the flowed discrete series 
$\hat{\cD}^{+ \,(p)}_j$ in \cite{MO}).
$\cW_{j,m,p}$ is  obviously  a subspace of $\cF_{j,m,p}$ (and they are not
isomorphic). 
Moreover, at least under the restriction 
$\dsp \frac{1}{2}<j\leq \frac{k}{2}$ \eqn{j-restriction},
we can show that $\cW_{j,m,p}$ can be  identified with some
(reducible, in general) $\widehat{SL}(2;\br)$-module,  since 
we have no singular vectors in the corresponding Verma module
(except for the Fock vacua themselves).  
It is easy to see that 
\begin{equation}
 \prod_{i}\cL^{(p)}_{-n_i} \ket{j,m,p} \in \bigoplus_{r\in \bsz} 
  \,\cW_{j, m+\frac{r}{p}, p}.
\label{lcw}
\end{equation}
Therefore  we can successfully realize the actions of DDF operators
$\cL_n^{(p)}$  within the (reducible) representations 
corresponding to $\dsp \bigoplus_{r\in \bsz} \,\cW_{j, m+\frac{r}{p}, p}$.

For $p>0$, the essentially same argument works by introducing 
the ``inverse Wakimoto representation'';
\begin{equation}
\left\{
\begin{array}{l}
 \dsp j^3 =\tilde{\beta}\tilde{\gamma} 
 -\sqrt{\frac{k}{2}}\partial \varphi\\
 \dsp j^+ = \tilde{\beta} \\
 \dsp j^- = \tilde{\beta}\tilde{\gamma}^2 
 -\sqrt{2k}\tilde{\gamma}\tilde{\varphi}-(k+2) \partial \tilde{\gamma} ,
 \end{array}
\right.
\end{equation}
or more explicitly,
\begin{equation}
\left\{
\begin{array}{l}
 \dsp \tilde{\varphi} =\rho - \sqrt{\frac{2k}{(k+2)}} Y^+ \\
 \dsp \tilde{\beta} = \left(-\sqrt{\frac{k+2}{2}}i\partial Y^1 
   + \sqrt{\frac{k}{2}}\partial \rho\right)\,e^{-\frac{2}{\sqrt{k+2}}iY^+}\\
 \dsp \tilde{\gamma} = e^{\frac{2}{\sqrt{k+2}}iY^+} ~~. 
\end{array}
\right.
\end{equation}

Consequently our choice of the representations of $\widehat{SL}(2;\br)$
is much larger than that of \cite{MO}
for the cases of $j \in \br$, although the enlargement 
of the Hilbert space by the spectral flow in \cite{MO} 
is incorporated into our setup as the discrete light-cone momentum $p$. 
By construction our physical Hilbert space also contains no ghosts.
In the analysis in \cite{MO} $m$ must take discrete values related with 
a fixed $j$ (belonging to the discrete series 
of $\widehat{SL}(2;\br)$ transformed by the spectral flow); $j-m \in \bz$. 
On the other hand, in our analysis {\em $m$ is arbitrary and independent of
$j$\/} as long as it satisfies the on-shell condition. This is natural 
from our starting point: the $\sigma$-model \eqn{ads3 lagrangian}
rather than the abstract representation theory of affine Lie algebra, 
and thus $j$ and $m$ (and of course $p$, too) should correspond to  
independent momenta along the different directions.

Furthermore, $\hat{\cD}^{+\,(p)}_j$ and 
$\hat{\cD}^{-\, (p-1)}_{\frac{k+2}{2}-j}$ are identified in \cite{MO},
since they are equivalent as an irreducible representation
of $\widehat{SL}(2;\br)$. Nevertheless they should be distinguished 
from our viewpoint, because they possess the different light-cone momenta $p$.
Especially, the standard discrete series $\hat{\cD}^{+}_j$ (lowest weight
representations), $\hat{\cD}^{-}_j$ (highest weight representations) 
are realized in the sectors $p=-1$, $p=1$ respectively, 
since the sectors $p=0$, $j\in \br$ are excluded  in our analysis.

We also mention that the above unitarity condition \eqn{j-restriction}
is analogous to the result given in \cite{EGP,MO}, but it is a 
slightly stronger condition. 
The unitarity bound proposed in \cite{EGP} reads 
$\dsp \frac{1}{2}<j <\frac{k+2}{2}$ and the one given in \cite{MO}
reads $\dsp \frac{1}{2}<j < \frac{k+1}{2}$
in our convention\footnote
   {The unitarity bound  
   for \cite{MO} is stronger than that of \cite{EGP} due to
    the identification $\hat{\cD}^{+\,(p)}_j
   =\hat{\cD}^{-\, (p-1)}_{\frac{k+2}{2}-j}$ 
    mentioned above. Moreover, the same range of $j$ was proposed in 
    a different context \cite{GK} by requiring good behaviors of
    the two point functions of the non-normalizable primary operators.
    Such two point functions nicely behave, too, under our constraints 
    \eqn{j-restriction}, since it is more stringent than that of \cite{GK}.}.  
This disagreement originates from the different choices 
of the representations mentioned above. 
In fact, if one choose to restrict the value of $m$ as $j-m \in \bz$
when solving the on-shell condition, one can show the no-ghost theorem
under the assumption $\dsp \frac{1}{2}< j < \frac{k+2}{2}$ same as \cite{EGP}
rather than \eqn{j-restriction}\footnote
      {To be precise, under this restriction $j-m\in \bz$ we must 
       take $\cL_n$ as the DDF operators rather than
       $\cL_n^{(p)} =p\cL_{n/p}+\cdots$, (recall \eqn{lcw}) 
       and hence the unitarity here means that $\cL_n$-descendants
       should not include any negative norm states.}.
It is not yet clear whether our choice of the momenta $m$, independent
of $j$, is completely valid even in the rigid treatment 
as an interacting theory, or nothing but an artifact originating 
from  the free field approximation.  
However, we again emphasize  that our setup of physical Hilbert space
admits the whole actions of DDF operators $\{\cL^{(p)}_n\}$ 
(and necessarily also with the space-time Virasoro algebra $\{\cL_n\}$). 
This fact is found to be consistent with the several results
about the long string sectors given by the light-cone gauge approach 
\cite{YZ,SW,HS2}, as we will comment in the next section. 
In fact, one can readily find that our choice of $m$ 
so as to be {\em independent of $j$\/} 
is crucial in order to ensure the equivalence with the spectrum 
in the light-cone gauge after solving the on-shell condition.   
In this sense we believe that our physical spectrum is valid at least 
for the long string configuration near the boundary,
which is well described  by the light-cone gauge approach.
In order to justify this spectrum beyond the near boundary region
we will have  to carry out a further analysis  
with the Liouville interaction term treated more precisely.

%%%%%%%%%%%%%%%%%%%%%%%%%%%%%%%%%%%%%%%%%%%%%%%%%%%%%%%%%%%%%%%%%%%%%

%One crucial difference from the results 
%of \cite{MO} is the spectrum of $m$-value. 
%In the analysis in \cite{MO} $m$ must take discrete values for 
%a fixed $j$ (belonging to the discrete series of $\widehat{SL}(2;\br)$ 
%with spectral flows). 
%On the other hand, in our analysis 
%$m$ can take a continuous value irrespective of $j$ as long as 
%it satisfies the on-shell condition. 
%This difference is originating 
%from the set up of Hilbert space: Our Hilbert space is 
%the Fock space of free fields on which  the DDF operators
%$\cA^{(p)}_n$, $\tilde{\cL}^{(p)}_n$ can act, but 
%that of \cite{EGP,MO} is defined as an $\widehat{SL}(2;\br)$-module,
%in which the meaning of  $\cA^{(p)}_n$, $\tilde{\cL}^{(p)}_n$ is 
%not clear. 

%%%%%%%%%%%%%%%%%%%%%%%%%%%%%%%%%%%%%%%%%%%%%%%%%%%%%%%%%%%%%%%%%%%%%%%

The extension of the above arguments to superstring examples
is not so difficult and we do not present it here. 
We instead focus on the spectrum of on-shell chiral primaries
of superstring on $AdS_3\times S^1 \times \cN /U(1)$ in the next section.

~

%%%%%%%%%%%%%%%%%%%%%%%%%%%%%%%%%%%%%%%%%%%%%%%%%%%%%%%%%%%%%%%%%%%%%%%%%%%%
\section{Chiral Primaries and Spectral Flow}

\hspace*{4.5mm}

In this section we further study  the spectrum 
in the superstring cases. 
We especially investigate an important class of observables:
chiral primaries. In other words, we shall concentrate on the 
topological sector of superstring vacua on 
$AdS_3\times S^1 \times \cN /U(1)$ \cite{Sugawara}. 
They are significant from the perspective of $AdS_3/CFT_2$-duality. 
Although we have not yet achieved the complete understanding 
of this duality, the study of their  spectrum  
will certainly clarify some aspects of it. 

Through this section we only deal with the left moving parts of 
objects, and it is easy to complete our discussions by taking
also the right movers.

\hspace*{4.5mm}

\subsection{Background with Space-time $N=4$ SUSY}

\hspace*{4.5mm}

We first discuss the most familiar
superstring vacua with space-time $N=4$ SUSY;
\begin{equation}
 AdS_3 \times S^3 \times T^4 \cong  AdS_3 \times S^1 \times
SU(2)/U(1)\times T^4, 
\end{equation} 
where $SU(2)/U(1)$ means that the Kazama-Suzuki model \cite{KazS} 
for this coset with $\dsp c=3-\frac{6}{N}$ ($N-2$ is equal to the level of 
(bosonic) $SU(2)$-WZW model describing the $S^3$ sector),
which is identified with the $N=2$ minimal model of $A_{N-1}$ type
and we denote it by $M_N$ from now on.
The criticality condition gives $k=N$, 
and this background is regarded as the near horizon limit of 
$NS5/NS1$ system, as is well-known\cite{AdS,GKS}.

Let $\ket{\Phi_l}$ ($l = 0,1,\ldots, N-2$) be the chiral primary states 
in the $M_N$ sector with $\dsp h_{\cN}=\frac{q_{\cN}}{2}= \frac{l}{2N}$.
We must look for the chiral primary states in the total system 
by tensoring $\ket{\Phi_l}$ with suitable vertex operators in 
the $AdS_3 \times S^1$ sector.
In the notation of previous section, 
namely,
\begin{equation}
\ket{j,m,p,q}\equiv \lim_{z\rightarrow 0}\,
 e^{ \left( \frac{\sqrt{k}}{2} p 
 - \frac{2m}{\sqrt{k}} \right) i X^+ 
 + \frac{\sqrt{k}}{2}p i X^- - \sqrt{\frac{2}{k}}j \rho 
 - \sqrt{\frac{2}{k}}q iX^2 }\ket{0},
\end{equation}
the possible candidates for the desired vertices are written as follows;
%(we only consider the left-mover);
\begin{equation}
\ket{j,j,p,j-\frac{k}{2}p}, ~~~
\Psi^+_{-1/2}\ket{j,-(j-1),p,-(j-1)-\frac{k}{2}p}.
\end{equation}
They are primary states with respect to $T(z)$, $G(z)$ and 
also satisfy
\begin{equation}
G^+_{-1/2}\ket{j,j,p,j-\frac{k}{2}p}
=G^+_{-1/2}\Psi^+_{-1/2}\ket{j,-(j-1),p,-(j-1)-\frac{k}{2}p}=0.
\end{equation}
First we consider $\ket{j,j,p,j-\frac{k}{2}p}\otimes\ket{\Phi_l}\otimes 
c_1e^{-\phi}\ket{0}_{\msc{gh}} $.
The on-shell condition leads to
\begin{equation}
j=\frac{N-l}{2}.
\label{cp1}
\end{equation}
(The GSO condition is automatically satisfied, since we have 
the relation $\dsp h=\frac{Q}{2}$.) 
Similarly, for the second candidate 
$\Psi^+_{-1/2}\ket{j,-(j-1),p,-(j-1)-\frac{k}{2}p}\otimes \ket{\Phi_l}\otimes 
c_1e^{- \phi}\ket{0}_{\msc{gh}}$ 
we can solve the on-shell condition and obtain 
\begin{equation}
j=1 + \frac{l}{2}.
\label{cp2}
\end{equation}

From now on we denote the first type 
of chiral primary \eqn{cp1} as $\ket{l,p\,;1}$
and the second type \eqn{cp2} as $\ket{l,p\,;2}$. 
Namely, we set
\begin{eqnarray}
\ket{l,p\,;1}&:=& \ket{\frac{N-l}{2}, \frac{N-l}{2}, p , -\frac{N(p-1)+l}{2} }
   \otimes \ket{\Phi_l} \otimes ce^{-\phi} \ket{0}_{\msc{gh}} \label{cp41}\\
\ket{l,p\,;2}&:=& \Psi^+_{-1/2}
 \ket{\frac{l}{2}+1, -\frac{l}{2}, p , -\frac{Np+l}{2} }
   \otimes \ket{\Phi_l} \otimes ce^{-\phi} \ket{0}_{\msc{gh}}~~. \label{cp42}
\end{eqnarray}
They are both normalizable and satisfy the unitarity condition 
\eqn{unitarity-ineq2}.

Remarkably one can find that \eqn{cp41}, \eqn{cp42} 
are also chiral primaries with respect to the space-time superconformal
algebra $\{\cL_n,~\cG^{\pm}_r, ~ \cJ_n\}$, as suggested in \cite{Sugawara}.
They satisfy
\begin{eqnarray}
\cL_0\ket{l,p\,;1} &=& \frac{1}{2}\cJ_0\ket{l,p\,;1} 
      = \frac{1}{2}\{l+N(p-1)\}\ket{l,p\,;1}  \label{Jlp1} \\
\cL_0\ket{l,p\,;2} &=& \frac{1}{2}\cJ_0\ket{l,p\,;2} 
      = \frac{1}{2}(l+Np)\ket{l,p\,;2} ~~.   \label{Jlp2}
\end{eqnarray}   
Since the light-cone momentum $p$ can now  take an arbitrary integer, 
we have infinite number of on-shell chiral primaries.
All of them have the same $U(1)_R$ charge in the sense of world-sheet 
because of the the on-shell condition. But they have the different 
$U(1)_R$ charges with respect to the space-time conformal algebra.

%%%%%%%%%%%%%%%%%%%%%%%%%%%%%%%%%%%%%%%%%%%%%%%%%%%%%%%%%%%%%%%%%%%%%%
Let us study the action of the spectral flow on these states. 
A natural extension of the spectral flow \eqn{sflow1} to the superstring 
case is given by
\begin{eqnarray}
  {\rm U}_p  X^0(z) {\rm U}_p^{-1} &=&  X^0(z) - p \sqrt{\frac{k}{2}}i \ln z
    \nonumber\\ 
 {\rm U}_p  X^2(z) {\rm U}_p^{-1} &=&  X^2(z) + p \sqrt{\frac{k}{2}}i \ln z,
\label{sflow3}
\end{eqnarray}
and the other fields should remain unchanged by the spectral flow.
The total $\widehat{SL}(2;\br)$-currents are transformed by them as follows;
\begin{eqnarray}
  {\rm U}_p J_n^3   {\rm U}_p^{-1} &=& J_n^3 +\frac{kp}{2}\delta_{n,0} 
      \nonumber\\
  {\rm U}_p J_n^\pm {\rm U}_p^{-1} &=& J_{n \mp p}^\pm~~.
\label{sflow4}
\end{eqnarray}
We can show that  
\begin{equation}
{\rm U}_r\ket{l,p\,;i} = \ket{l,p+r\,;i}.
\label{sflow-cp}
\end{equation}
Namely, the spectral flow maps  an on-shell chiral primary 
to another on-shell chiral primary. 
This is a general feature. In fact,
it is a straightforward calculation to check that 
\begin{eqnarray}
{\rm U}_p G^+(z){\rm U}_p^{-1}&=& G^+(z)  \\
{\rm U}_pQ_{BRST}{\rm U}_p^{-1} &=& Q_{BRST} -
\lim_{z\rightarrow 0}\left[pc\sqrt{\frac{k}{2}}i\partial (X^0+X^2) 
+ p\eta e^{\phi}\sqrt{\frac{k}{2}}(\Psi^0+\Psi^2)\right] .
\label{sflow-brst}
\end{eqnarray}
Because the correction term in \eqn{sflow-brst} vanishes 
when acting on an arbitrary on-shell chiral primaries, 
the spectral flow ${\rm U}_p$ preserves the on-shell condition 
in the space of chiral primaries. One should remark that 
${\rm U}_p$ does {\em not\/} transform all  the physical states
among themselves.
Indeed the physical states which are not the chiral
primaries are transformed into off-shell states by ${\rm U}_p$.

Turning our attention to the space-time conformal algebra,
we obtain
\begin{eqnarray}
 \{{\cal G}^-_{\frac{1}{2}}, {\cal G}^+_{-\frac{1}{2}}\}
 &=& {\cal L}_0 - \frac{1}{2}{\cal J}_0 \nonumber\\
% &=& - \oint \sqrt{\frac{k}{2}} i \partial X^0 
%   - \frac{1}{2}  \oint \sqrt{2k} i \partial X^2 \nonumber\\
 &=& - \oint \sqrt{\frac{k}{2}} i \partial (X^0 + X^2)~~,
\end{eqnarray}
and this identity is unchanged by the spectral flow.
This means that the spectral flows are closed in the space of
the {\em space-time\/} chiral primaries, which is consistent with 
the above observation.

%%%%%%%%%%%%%%%%%%%%%%%%%%%%%%%%%%%%%%%%%%%%%%%%%%%%%%%%%%%%%%%%

~

Next we present some remarks from the view points of $AdS_3/CFT_2$-duality
and the long string theory given in \cite{SW}.
Although our understanding of them by string theory is not yet 
complete, we believe the following remarks are useful to 
clarify some important aspects of them.

Tracing back to the procedure of our field redefinitions, one  
can find that $\ket{l,p=1\, ;1}$ can be identified with 
the space-time chiral primary states given in \cite{HS1} 
(see also \cite{KLL})\footnote
   {They correspond to $\ket{\om^0_{l/2}}$ in the notation of
    \cite{HS1}, which contain the trivial cohomology in the $T^4$
    sector. We can also consider more general {\em space-time\/}
   chiral primary states that contain higher cohomologies 
    of $T^4$, as given in  \cite{HS1}. But they are not chiral primaries
    in the sense of world-sheet.}.  
It has the following structure
\begin{equation}
\ket{l,1\,;1} = \lim_{z\rightarrow 0}\cO_{l}(z)\ket{0, 1\,;1 }, \\
\end{equation}
where 
\begin{equation}
\cO_l(z) := 
V_{-\frac{l}{2},-\frac{l}{2}, 0,-\frac{l}{2}} \,\Phi_l(z) 
\equiv e^{\frac{l}{\sqrt{2N}}(iX^0+iX^1+iX^2+\rho)}\,\Phi_l(z)
\end{equation} 
naturally corresponds to  the chiral operator in the light-cone
gauge formalism of the long string theory given in \cite{SW}.  
$\ket{0, 1\,;1 }
\equiv e^{-\sqrt{\frac{N}{2}}(iX^1+\rho)}\, ce^{-\phi}\ket{0}$ 
has the maximal $j$-value $\dsp j=N/2\equiv k/2$
and  is the same as the ``space-time vacuum'' (or ``long string vacuum'') 
presented in \cite{HS1}. It satisfies 
\begin{eqnarray}
\cL_n \ket{0, 1\,;1} &=&0 ,~~~(\forall n\geq -1) , \label{Lvac} \\
\cG^{\pm}_r \ket{0, 1\,;1} &=&0 , ~~~ (\forall r\geq -1/2) , \label{Gvac} \\
\cJ_n \ket{0, 1\,;1} &=& 0 ,~~~(\forall n\geq -1) , \label{Jvac} \\
-\frac{\sqrt{k}}{2}\oint i\partial X^+ \ket{0, 1\,;1} 
&=& \ket{0, 1\,;1}~, \label{Wvac}
\end{eqnarray}
where the last line simply means that 
$\dsp p\equiv \oint \gamma^{-1}\partial \gamma =1 $, and 
these identities hold up to BRST-exact terms.

As discussed in \cite{SW} (see also \cite{KutS2}), 
the chiral operator $\cO_l$ corresponds
to a non-normalizable state. Its  wave function 
is divergent near the boundary, where the Coulomb branch CFT
is weakly coupled. On the other hand, thanks to the existence of
$\ket{0, 1\,;1}$, the chiral primary state $\ket{l,1;1}$ itself
becomes normalizable state vanishing exponentially
at large $\rho$, as expected from the observation about 
the Higgs branch tube in \cite{SW}.
In this sense the interpretation of $\ket{0, 1\,;1}$ as the long 
string vacuum may be natural.  

It may be also useful to define explicitly  
the {\em space-time\/} chiral primary operator $\dsp \hat{\cO}_l(x)\equiv
\sum_{n}\frac{\hat{\cO}_{l,n}}{x^{n+\frac{l}{2}}}$ 
by introducing the vertex operators
\begin{equation}
\hat{\cO}_{l,n} := \oint\, V_{-\frac{l}{2},n,0,-\frac{l}{2}}
               \Phi_l(\Psi^0+\Psi^1)e^{-\phi}~, 
\label{cp-mode}
\end{equation}
where $n$ runs over all (half-)integers 
if $\dsp \frac{l}{2}$ is an (half-)integer.
$\hat{\cO}_l(x)$ actually behaves as a chiral primary operator with respect 
to the space-time SCA. For example, we obtain
\begin{equation}
[\cL_m, \hat{\cO}_{l,n}] = \left\{(\frac{l}{2}-1)m-n \right\} 
\hat{\cO}_{l,m+n},
\end{equation}
which means that $\hat{\cO}_l(x)$ is a primary operator with conformal 
weight $\dsp h=\frac{l}{2}$. We can further show that 
\begin{equation}
 \hat{\cO}_{l,n}\ket{0,1\,;1}=0 ,~~~ (\forall n>-\frac{l}{2})
\end{equation} 
and also obtain the ``operator-state correspondence''
\begin{equation}
\hat{\cO}_{l,-\frac{l}{2}}\ket{0,1\,;1}= \ket{l,1\,;1} 
\label{os}
\end{equation}
(up to the picture changing and an overall constant).

For $p>1$ we can consider more general chiral primaries 
with the higher space-time $U(1)_R$-charges
\begin{equation}
\ket{l,p\,;1}= {\rm U}_{p-1}\ket{l,1\,;1}
= \cO_l(0)\ket{0 , p\,;1} 
=\hat{\cO}_{l, -\frac{l}{2}}\ket{0 , p\,;1} ~~.
\end{equation}
As we observed above \eqn{Jlp1}, the spectrum of $\cJ_0$ charge 
is $l+N(p-1)$, $l=0,1,\ldots, N-2$, $p\geq 1$.

We have not yet known the suitable interpretation  
of the ``graviton-like'' chiral primary states $\ket{l,p\,;2}$
in the context of $AdS_3/CFT_2$ correspondence. 
We only point out that they do not seem to have 
the forms such as \eqn{os}, and so it might be plausible to suppose
that they do not have any counterparts in the boundary theory,
as long as  our identification of $\ket{0,1\,;1}$ 
with the space-time vacuum is justified.
In any case we will need a further analysis to give a more definite  
statement about this problem.

The following aspect may be worthwhile to point out.
Here the chiral primaries with the higher space-time $U(1)_R$-charges 
appear in the sector with higher light-cone momenta $p$
(or by taking account of the degrees of freedom of spectral flows).
On the other hand, in the analysis of \cite{HS2}, they correspond 
to the $\bz_p$-twisted sector of the symmetric orbifold theory,
which describes the sector of long string with the ``length'' $p$
as in the Matrix string theory \cite{MST}.
It suggests a remarkable correspondence between the spectral flow
in the covariant gauge formalism  and the twisted sector of the
symmetric orbifold in the light-cone gauge formalism \cite{YZ,SW}.

To address the precise correspondence between them we should work on
the {\em second quantized\/} framework. It is quite reasonable from the
viewpoints of $AdS_3/CFT_2$ correspondence, 
since the boundary CFT should also contain 
multi-particle excitations. 
We shall now focus on the physical states 
with positive energies, which should have 
the light-cone momenta $p\geq 0$ as we found in section 3. 
The physical Hilbert space of the first quantized  string states,
which was studied in our previous analyses, is then 
decomposed to $\dsp \cH = \bigoplus_{p\geq 0}\cH_p$,
where $\cH_p$ denotes the sector with the light-cone momentum $p\geq 0$.
The Hilbert space in the (free)
second quantized theory can be roughly written as 
\begin{equation}
\hat{\cH} = \bigoplus_{n=0}^{\infty}\, (\cH)^{\otimes n}.
\label{second}
\end{equation}
(To be precise, we must make some (anti-)symmetrization to 
assure the correct statistics in this and 
the expressions given below.) 
The second quantized space $\hat{\cH}$
has a natural decomposition with respect to the total light-cone momentum
\begin{equation}
\hat{\cH} = \bigoplus_{p \geq 0} \hat{\cH}_p, ~~~ 
%\hat{\cH}_p =\bigotimes_{\scriptstyle{\sum_i p_i =p}}\cH_{p_i}.
\label{hhatp}
\end{equation}
Obviously $\hat{\cH}_p$ is decomposed to the subspaces of the forms
\begin{equation}
 \cH_{p_1}\otimes \cH_{p_2} \otimes \cdots \otimes \cH_{p_n} 
\otimes \cH_0 \otimes \cH_0 \otimes \cdots ,~~~ 
(p_1\geq p_2 \geq \ldots \geq p_n \geq 1, ~ \sum_{i=1}^n p_i=p ) ~.
\label{mls} 
\end{equation}  
The $p=0$ Hilbert space $\cH_0$ only contains tachyons,  which are 
eliminated by the GSO projection, as was already shown\footnote
      {The fact that the physical Hilbert space 
       of  ``short string sector'' $\cH_0$ is vacant 
       is not a contradiction. It rather means that only 
       the non-normalizable  physical {\em operators\/} can appear 
      and the operator-state correspondence fails 
      in the short string sector as suggested in \cite{DORT}.}. 
Therefore we can neglect the $\cH_0$ factors, 
and can explicitly write down
\begin{equation}
\hat{\cH}_p = \bigoplus_{n=1}^{p}\, \bigoplus_{\sum_{i=1}^n p_i=p} 
  (\otimes_{i=1}^n \cH_{p_i})~~.
\label{hhatp2}
\end{equation}

Now let us consider the system of  $Q_5$ NS5 and $Q_1$ NS1.
The NS5 charge $Q_5(\equiv N)$ appears in the world-sheet action
of $AdS_3$-string theory, but $Q_1$ does not. It only appears 
in the string coupling, which is stable under the near horizon limit,
and hence we cannot find this effect in the first quantized theory.
However, in the second quantized theory, it is quite natural 
to identify the NS1 charge $Q_1$ with the total light-cone momentum 
%(or the ``total winding number of the multiple long strings'') 
$p$ in the expression of $\hat{\cH}_p$.
Hence we propose that the physical Hilbert space of this NS5-NS1
system should be defined as $\hat{\cH}_{Q_1}$. Notice that it has  
the structure characterized by the various partitions
$\dsp \{p_i(\geq 1)\,;\, \sum_i p_i=Q_1\}$ which is 
consistent with the expected  correspondence to the
symmetric orbifold theory.
Clearly this system can be decomposed to 
the subsystems of various long strings with
the ``lengths'' (or ``windings'') $p_i$ ($1 \leq p_i\leq Q_1$
$\dsp \sum_i p_i=Q_1$), 
as observed in \cite{SW,KS}.

It is interesting to present the spectrum of the ``single-particle
chiral primaries'' in this framework. Let $p \leq Q_1$ be a positive
integer. We obtain the required states as  
\begin{equation}
\ket{l,p\,;1}\otimes 
\underbrace{\ket{0,1\,;1}\otimes \cdots \otimes\ket{0,1\,;1}}
_{\msc{$Q_1-p$-times}} ~,
\end{equation} 
which satisfies  $\dsp \cL_0=\frac{1}{2}\cJ_0= \frac{l+Q_5(p-1)}{2}$ 
as expected.
Since $l$, $p$ run over  the ranges $0\leq l \leq Q_5-2$, 
$0\leq p \leq Q_1$,
this spectrum is completely in agreement  with the result of  \cite{HS2}, 
in which the (multiple) long string CFT was
analyzed using the symmetric orbifold theory. 
%(with the symmetric group $S_{Q_1}$).
Quite remarkably, this has the upper bound $\dsp \sim \frac{Q_1Q_5}{2}$
which is expected from the $AdS_3/CFT_2$-duality
\cite{MS}, as already commented in \cite{HS2}.

Notice that there are the missing states corresponding to the $\cJ_0$-charge
$\dsp Q_5 p-1$ $(1\leq p \leq Q_1)$. They  are (formally) written 
as 
\begin{equation}
\ket{Q_5-1,p\,;1} \otimes \ket{0,1\,;1} \otimes \cdots 
= \cO_{Q_5-1}(0)\ket{0,p\,;1}  \otimes \ket{0,1\,;1} \otimes \cdots ~~, 
\label{missing}
\end{equation}
and ``$\cO_{Q_5-1}(z)$'' is no other than the missing chiral operator 
discussed in \cite{SW}, which should correspond to the cohomology
with a delta function support at the small instanton singularity.
In this sense these missing states \eqn{missing}
are supposed to be the natural generalizations to the cases of $1<p\leq Q_1$ 
of the one discussed in  \cite{SW}, in which only the $p=1$ sector
was treated.

The above observation implies  that the first quantized Hilbert space
$\cH_p$ ($p\leq Q_1$) precisely corresponds to the $\bz_p$-twisted
sector in the $S_{Q_1}$-orbifold theory 
%(corresponding to the Young
%tableau $(p,\underbrace{1,1,\ldots,1}_{\msc{$Q_1-p$ times}})$),
as we already suggested.
The relation
\begin{equation}
\cL_n = \frac{1}{p}\cL_{pn}^{(p)} + \frac{Q_5}{4}\left(p-\frac{1}{p}\right)
 \delta_{n,0} 
\label{Lnp}
\end{equation}
indeed confirms this identification.
On the Hilbert space $\cH_p$, 
the space-time Virasoro algebra $\cL_n$ has the central charge $c=6pQ_5$,
and the DDF operators $\cL_n^{(p)}$ generate the Virasoro algebra 
with $c=6Q_5$. This relation \eqn{Lnp} is the same as the well-known 
formula to define the conformal algebra describing the $\bz_p$-twisted sector
of the symmetric orbifold. 
It is easy to define the tensor product
representation of space-time conformal algebra
with $c=6Q_1Q_5$ on the second quantized Hilbert space 
$\dsp \hat{\cH}_{Q_1}\equiv \bigoplus_{\sum_i p_i=Q_1}\, (\otimes \cH_{p_i})$
including the conformal invariant vacuum
\begin{equation}
  \underbrace{\ket{0,1\,;1}\otimes \ket{0,1\,;1}\otimes
\cdots\otimes\ket{0,1\,;1}}_{Q_1 \msc{times}} \, \in \cH_1^{\otimes Q_1} ~~. 
\label{civac}
\end{equation}
Such a correspondence, which was essentially suggested in 
\cite{YZ,HS2}, is quite expected from our standpoint 
as the discrete light-cone theory fitted to the spirit of 
Matrix string \cite{DLCQ,MST}. 
Recall that our setup of physical Hilbert space in section 3 allows 
the action of $\cL^{(p)}_n =p\cL_{n/p}+ \cdots$,
and moreover we must impose the ``level matching condition''
$\cL_0^{(p)}-\overline{\cL^{(p)}_0}
\in p \bz$ onto the Hilbert space $\cH_p$. 
These facts are crucial to establish the above correspondence 
to the symmetric orbifold.

One should keep in mind the following fact:
one can also construct the representation with $c=6Q_1Q_5$ 
on the first quantized Hilbert space $\cH_{Q_1}$ that 
is the subspace of $\hat{\cH}_{Q_1}$ describing the {\em single\/} 
long string with the maximal length $Q_1$. 
{\em However, $\cH_{Q_1}$ cannot include the conformal invariant
vacuum.} Recall that $\cL_0\ket{0,p\,;1}\neq 0$, unless $p=1$.
More precisely speaking, we can show that, 
in our setup of the first quantized Hilbert space
the BRST-invariant state with the properties
\eqn{Lvac}, \eqn{Gvac}, \eqn{Jvac} and non-zero $p$ is possible 
only if $p=1$, and the solution is  unique
(up to BRST exact terms and an overall normalization),
$\ket{0,1\,;1}$, as suggested in \cite{HS1,HS2}.
%from an estimation of  physical degrees of freedom.
This fact leads us to the only one possibility of the 
conformal invariant vacuum \eqn{civac}.
The large Hagedorn density suited for $c=6Q_1Q_5$, 
which may reproduce the correct entropy formula of black-hole,
should be attached to $\hat{\cH}_{Q_1}$, {\em not\/} to $\cH_{Q_1}$,
since $\cH_{Q_1}$ does not include the vacuum state such that $\cL_0=0$
(see the discussions given in  \cite{KutS2,Carlip}).  
%In other words, one should recall that 
%the spectrum  generating algebra for the first quantized space
%$\cH_p$ should be $\{\cL^{(p)}_n\}$ (not $\{\cL_n\}$). 

%Recall our assertion that 
%the bound string states ($j\in \br$) with positive energy
%should have $p>0$ based on the unitarity argument
%(likewise, the bound string states with negative energy must 
%have $p<0$). This is consistent with the reinterpretation
%of the momentum $p$ as the number of strings glued one after another. 
%Especially, it is crucial for this interpretation
%that all the sectors $j\in \br$, $p=0$  are excluded.

~

%%%%%%%%%%%%%%%%%%%%%%%%%%%%%%%%%%%%%%%%%%%%%%%%%%%%%%%%%%%%%%%%%%%%%%%

\subsection{Background with Space-time $N=2$ SUSY}

\hspace*{4.5mm}

In principle it is not difficult to generalize the above analysis on 
chiral primaries to more general superstring vacua with 
space-time $N=2$ SUSY \cite{GRBL}. 

We first give a rather generic argument.
Consider superstring theory on $AdS_3\times S^1\times \cN /U(1)$,
where $\cN /U(1)$ is an arbitrary $N=2$ SCFT of center $9-6/k$.
%and was denoted by ${\cal N}/U(1)$ in the previous section.
As in the $N=4$ case, we can construct two series of 
chiral primary states from  chiral primaries $V_j$
of conformal weight $j/2k$ in the $\cN /U(1)$-sector:
\begin{eqnarray}
\ket{j,p\,;1}&:=& 
  \ket{\frac{k-j}{2},\frac{k-j}{2}, p, -\frac{k(p-1)+j}{2}}\otimes
  \ket{V_j} \otimes
  ce^{-\phi} \ket{0}_{\rm gh} \\
\ket{j,p\,;2}&:=& \Psi^+_{-1/2}
  \ket{\frac{j+2}{2},-\frac{j}{2}, p, -\frac{kp+j}{2}}\otimes
  \ket{V_j} \otimes
  ce^{-\phi} \ket{0}_{\rm gh}
\end{eqnarray}
They have the light-cone momentum  $p$ and the following conformal weight:
\begin{eqnarray}
    {\cal L}_0\ket{j,p\,;1}
   =\frac{1}{2}{\cal J}_0\ket{j,p\,;1}
 &=&\frac{j+k(p-1)}{2}\ket{j,p\,;1} \nonumber \\
    {\cal L}_0\ket{j,p\,;2}
   =\frac{1}{2}{\cal J}_0\ket{j,p\,;2}
 &=&\frac{j+kp}{2}\ket{j,p\,;2}
\end{eqnarray}

Note that, if we take as $\cN /U(1)$ an arbitrary $N\!=\!2$ SCFT
of center $9-6/k$, the conformal weight $h=j/2k$ of $V_j$
runs within the range $0\le h \le 3-2/k$.
However, it is only if $0\le h \le 1/2-1/k$ that
the chiral primary states are in the spectrum
allowed from unitarity and normalizability.

Let us consider a specific example.
Take as $\cN/U(1)$ the $N=2$ minimal model
which we denote by $M_N$ as before.
It was proposed in \cite{GiveonKP}
that the superstring theory on this background
is marginally equivalent to the non-critical superstring theory 
\cite{KutS1} which is  holographically dual to the decoupled theory 
based on the $A_{N-1}$-singular $CY_4$. 
In this case the criticality condition leads to $\dsp k=\frac{N}{N+1}$.
%It is straightforward to replace the $M_N$ sector with  a more general 
%unitary $N=2$ $SCFT_2$. 
%Notice that, since here $k<1$ holds, 
%we have no normalizable physical states compatible with the unitarity 
%constraints \eqn{unitarity-ineq2},
%according to our previous analysis.

Let $\ket{\Phi_l}$ ($l = 0,1,\ldots, N-2$) be again the chiral primary states 
of weight $l/2N$ in the $M_N$ sector.
We obtain the following chiral primaries;
\begin{eqnarray}
\ket{l,p\,;1}&:=& \left| \frac{N-l}{2(N+1)}, \frac{N-l}{2(N+1)}, 
     p , -\frac{N(p-1)+l}{2(N+1)} \right\rangle
   \otimes \ket{\Phi_l} \otimes ce^{-\phi} \ket{0}_{\rm gh} \label{cp21}\\
\ket{l,p\,;2}&:=& \Psi^+_{-1/2}
 \left| \frac{l}{2(N+1)}+1, -\frac{l}{2(N+1)}, p , 
     -\frac{Np+l}{2(N+1)} \right\rangle
   \otimes \ket{\Phi_l} \otimes ce^{-\phi} \ket{0}_{\rm gh}. \label{cp22}
\end{eqnarray}
In this way we have again infinite number of on-shell chiral primaries
possessing the following space-time $U(1)_R$ charges;
\begin{eqnarray}
{\cal J}_0 \ket{l,p\,;1}
&=&\left(\frac{l}{N+1}+\frac{N}{N+1}(p-1)\right)\ket{l,p\,;1}
\label{u1r1} \\
{\cal J}_0 \ket{l,p\,;2}
&=&\left(\frac{l}{N+1}+\frac{N}{N+1}p\right)\ket{l,p\,;2}
\label{u1r2} .
\end{eqnarray}

Unfortunately, $\ket{l,p\,;1}$ is non-normalizable and 
$\ket{l,p\,;2}$ does not satisfy the unitarity constraints 
\eqn{unitarity-ineq2}. Hence we cannot consider the chiral primary
{\em states\/} within the physical Hilbert space.
We must only treat these chiral primaries as {\em operators\/}
and cannot expect the operator-state correspondence.
%with respect to the chiral primaries in this example. 
%The chiral primaries cannot create the usual string states 
%that should be normalizable and have no ghosts in
%their DDF descendants.  
%There is no suitable space-time vacuum such as the one for 
%the $N=4$ background $AdS_3\times S^3 \times T^4$. 
Nevertheless, they may be regarded as an important class of operators 
in the context of $AdS_3/CFT_2$-duality, 
or more general holographic dualities \cite{ABKS,GiveonKP}. 
In particular the non-normalizable chiral primaries $\ket{l,p\,;1}$ 
(``tachyon-like operators'') may be important, 
because they possess  the momentum structures
which can be regarded as natural generalizations of those of   
the scaling operators in the space-time conformal theory
proposed in \cite{GiveonKP}. 
%which should describe 
%the decoupled singular $CY_4$ compactification. 
Since the light-cone momentum $p$ runs over an infinite range,
we can obtain the infinite tower of space-time chiral operators  
for each of the chiral operators in the ``matter sector'' $\Phi_l$. 
This aspect may be interesting, since they look  like 
analogues of ``gravitational descendants'' in the theory 
of two dimensional gravity. 
We must make further studies to gain more precise insights
about these objects. In addition, the roles of  the graviton-like operators 
$\ket{l,p\,;2}$ are again unclear. More detailed argument for them will be 
surely important, although it is beyond the scope of this paper.

~

%%%%%%%%%%%%%%%%%%%%%%%%%%%%%%%%%%%%%%%%%%%%%%%%%%%%%%%%%%%%%%%%%%%%%%%%%

\section{Conclusions and Discussions}

\hspace*{4.5mm}

In this paper we have studied the spectrum
of the physical states in string theory
on $AdS_3$ based on a free field realization. We have found that 
the system is quite simply described as a linear dilaton theory
with a light-like compactification, which we called as 
``discrete light-cone Liouville theory''.  
Our key idea is to utilize the DDF operators according to
the traditional approach to string theory on flat backgrounds.
We have {\em two\/} independent sets of DDF operators; $\cA^{(p)}_n$,
$\tilde{\cL}^{(p)}_n$. 
This situation is similar to the non-critical 
string, although we started with the {\em critical\/} string on 
$AdS_3\times \cN$ background. In fact, we can easily find that 
a suitable linear combination of $\cA^{(p)}_n$ and
$\tilde{\cL}^{(p)}_n$ gives the ``longitudinal DDF operator''  
utilized in \cite{Brower}.

Regarding the system as a free theory with no Liouville interaction term 
(screening charge term), the physical spectrum contains only 
the principal series $\dsp j=\frac{1}{2}+is$ as asserted in 
\cite{Bars,Satoh,BDM}, and the physical states are 
generated by the DDF operators $\{\cA^{(p)}_n,\, \tilde{\cL}^{(p)}_n\}$
mentioned above.

However, once we turn on the Liouville potential, 
the story becomes rather non-trivial. 
In this interacting theory the translational invariance along 
the Liouville direction is broken. The physical Hilbert space
is expected to be spanned only by the $\widehat{SL}(2;\br)$ currents,
rather than the whole oscillators of the string coordinates
$\rho, \, Y^+, \, Y^-$, because the interaction term 
commutes only with the $\widehat{SL}(2;\br)$ currents. 
In this situation the spectrum generating algebra  
becomes $\{\cL_n^{(p)}\}$ in place 
of $\{\cA^{(p)}_n,\, \tilde{\cL}^{(p)}_n\}$,
and the physical states possessing the  imaginary $\rho$-momenta
($j\in \br$) are also allowed. Physically they correspond to 
the bound string states of \cite{MO} 
that are trapped inside the $AdS_3$-space.

%%%%%%%%%%%%%%%%%%%%%%%%%%%%%%%%%%%%%%%%%%%%%%%%%%%%%%%%%%%%%%%%%%%%%%%%%%%

It may be worthwhile to mention that 
only the physical Hilbert space
as the interacting Liouville theory  
may be consistent with the microscopic evaluation of the  black hole entropy.
In the free system the DDF operators should be
$\{\cA^{(p)}_n,\,\tilde{\cL}_m^{(p)}\}$ and $\tilde{\cL}^{(p)}_m$ 
(which are identified with the Virasoro operators in $\cN$-sector under
the light-cone gauge) have a small central charge $\dsp c=23-\frac{6}{k}$. 
(An important discussion related to such a counting of physical states
was given in \cite{Carlip}.)
On the other hand, 
after turning on the Liouville potential we claimed that
the full Virasoro generators $\{\cL_n^{(p)}\}$ are well-defined
(and it is also crucial that $\{\cA^{(p)}_n\}$ should be discarded). 
Taking further account of the second quantized Hilbert space
%(describing the ``multiple long string system'')
they seem to generate sufficiently many physical states 
with the Hagedorn density that can reproduce the correct entropy.
%    \footnote
%   {We should here recall that the spectrum generating algebra is  
%    $\{\cL^{(p)}_n\}$, {\em not\/} $\{\cL_n\}$, and has the central charge
%     $c\sim 6k$ ({\em not\/} $6kp$).
%     This does not mean a failure of our discussion, since we are now 
%     working on the first quantized string theory.
%     To reproduce the large Hagedorn density suited for the central
%     charge $c \sim 6kp$, one must consider the system of multiple long
%     strings as in \cite{HS2}.}. 
We would like to study this problem in more detail elsewhere.

In the study of superstring examples,
we have presented the complete set of on-shell chiral primaries. 
There exist infinite number of such operators and 
the spectral flows naturally act on them. 
%Remarkably, we have the upper bound for space-time $U(1)_R$ charges
%of these chiral operators $\sim kp/2$ for each sector of fixed $p$. 
Moreover, to describe the well-known $Q_5(\equiv k)$ NS5 - $Q_1$ NS1 system 
we made use of the second quantized framework. The Hilbert space
of the multiple long string system was given as the form 
$\dsp \hat{\cH}_{Q_1}=\bigoplus_{\sum_i p_i=Q_1}\, 
(\otimes_i  \cH_{p_i})$, where $\cH_p$ denotes the first quantized 
physical Hilbert space of the sector with the light-cone momentum
$p~ (>0) $. This space reproduce the spectrum of chiral primaries 
same as that given by the symmetric orbifold theory \cite{HS2},
and among other things, we have  successfully obtained
the upper bound $\sim Q_1Q_5/2$ consistent 
with the prediction of $AdS_3/CFT_2$ correspondence \cite{MS}.

It may be also worth pointing out that our reformulation 
of superstring on $AdS_3\times S^1 \times \cN/U(1)$ 
%as the $N=2$ Liouville theory 
has the same field contents  as those of the non-critical string 
that is holographically dual to  a singular Calabi-Yau compactification
(especially, the cases of $CY_4$) proposed in \cite{GiveonKP}.
The only difference between these models is  the existence/absence
of the light-like compactification.  
In \cite{GiveonKP} it  was discussed that these two backgrounds 
can be interpolated by some marginal deformation. 
It may be an interesting problem to clarify 
the rigid correspondence between them. 
In particular, our analyses on general chiral operators in section 4 
will be readily generalized to the cases 
of such non-critical string theories.

~

%%%%%%%%%%%%%%%%%%%%%%%%%%%%%%%%%%%%%%%%%%%%%%%%%%%%%%%%%%%%%%%%%%%%%%%%%%%

\section*{Acknowledgement}

Y. S.  would like to thank I. Bars and Y. Satoh for helpful discussions.

The work of K. H. is supported in part by
Japan Society for Promotion of Science under
the Postdoctral Research Program ($\sharp 12$-$02721$).
The work of Y. S. is supported in part  by 
Grant-in-Aid for Encouragement of Young Scientists ($\sharp 11740142$) 
and also by Grant-in-Aid for Scientific Research on Priority Area 
($\sharp 707$) ``Supersymmetry and Unified Theory
of Elementary Particles", 
both from Japan Ministry of Education, Science, Sports and Culture.

%%%%%%%%%%%%%%%%%%%%%%%%%%%%%%%%%%%%%%%%%%%%%%%%%%%%%%%%%%%%%%%%%%%%%%%%%%


\newpage

\begin{thebibliography}{99}


\bibitem{sl2}
J. Balog, L. O'Raifeartaigh, P. Forgacs  and A. Wipf,
Nucl. Phys. {\bf B325} (1989) 225;
P. Petropoulos, 
Phys. Lett. {\bf B236} (1990) 151;
N. Mohammedi,
Int. J. of Mod. Phys. {\bf A5} (1990) 3201;
I. Bars and D. Nemeschansky, 
Nucl. Phys. {\bf B348}, 89 (1991);
%M. Henningson and S. Hwang, 
%Phys. Lett. {\bf B258}, (1991) 341; 
S. Hwang, 
Nucl. Phys. {\bf B354} (1991) 100.
%M. Henningson, S. Hwang, P. Roberts and B. Sundborg,
%Phys. Lett. {\bf B267}, 350 (1991);


\bibitem{Bars}
I. Bars, Phys. Rev. {\bf D53} (1996) 3308,
hep-th/9503205.

\bibitem{Satoh}
Y. Satoh, Nucl. Phys. {\bf B513}, 213 (1998),
hep-th/9705208.


\bibitem{AdS}
J. Maldacena,
%{\it``The Large $N$ Limit of Superconformal Field Theories
%and Supergravity''},
Adv. Theor. Math. Phys. {\bf 2} (1998) 231,
hep-th/9711200;
%\bibitem{GKP}
S. Gubser, I. Klebanov and A. Polyakov,
%{\it ``Gauge Theory Correlators from Non-Critical String Theory''},
Phys. Lett. {\bf B428} (1998) 105,
hep-th/9802109;
%\bibitem{Witten1}
E. Witten,
%{\it ``Anti De Sitter Space And Holography''},
Adv. Theor. Math. Phys. {\bf 2} (1998) 253,
hep-th/9802150;
%\bibitem{review}
O. Aharony, S. Gubser, J. Maldacena, H. Ooguri and Y. Oz,
hep-th/9905111 (for a review and a complete list of references).

\bibitem{GKS}
A. Giveon, D. Kutasov and N. Seiberg,
%{\it ``Comments on String Theory on $AdS_3$''},
Adv. Theor. Math. Phys. {\bf 2} (1998) 733,
hep-th/9806194.

\bibitem{DORT}
J. de Boer, H. Ooguri, H. Robins, and J. Tannenhauser,
%{\it ``String Theory on $AdS_3$''},
JHEP {\bf 9812} (1998) 026, hep-th/9812046.

\bibitem{KS}
D. Kutasov and N. Seiberg,
%{\it ``More Comments on String Theory on $AdS_3$''},
JHEP {\bf 9904} (1999) 008, hep-th/9903219.

\bibitem{BF}
D. Bernard and G. Felder,
Commun. Math. Phys. {\bf 127} 145 (1990).

\bibitem{EGP}
J. Evans, M. Gaverdiel and M. Perry,
Nucl. Phys. {\bf B535} (1998) 152,
hep-th/9806024.  

\bibitem{BDM}
I. Bars, C. Deliduman and D. Minic,
%``{\it String Theory on AdS_3 Revisited}'',
hep-th/9907087.

\bibitem{MO}
J. Maldacena and H. Ooguri,
hep-th/0001053.

\bibitem{BH}
J. Brown and  M. Henneaux, 
%{\it ``Central Charges In The Canonical Realization Of Asymptotic Symmetries:
%An Example from Three-Dimensional Gravity''},
Commun. Math. Phys. {\bf 104} (1986) 207.

\bibitem{Strominger}
A. Strominger,
%{\it ``Black Hole Entropy from Near-Horizon Microstates''},
JHEP {\bf 9802} (1998) 009,
hep-th/9712251.

\bibitem{MMS}
J. Maldacena, J. Michelson and A. Strominger,
JHEP {\bf 9902} (1999) 011, hep-th/9812073.

\bibitem{SW}
N. Seiberg and E. Witten,
%{\it ``The D1/D5 System and Singular CFT''},
JHEP {\bf 9904} (1999) 017, hep-th/9903224.

\bibitem{DDF}
E. Del Giudice, P. Di Vecchia and S. Fubini, 
Ann. Phys. {\bf 70} (1972) 378.

\bibitem{HS2}
K. Hosomichi and Y. Sugawara, 
JHEP {\bf 9907} (1999)027, hep-th/9905004.

\bibitem{Wakimoto}
M. Wakimoto,
Commun. Math. Phys. {\bf 104} (1986) 605.


\bibitem{FMS}
D. Friedan, E. Martinec and S. Shenker,
Nucl. Phys. {\bf B271} (1986) 93.

\bibitem{IK}
H. Ishikawa and M. Kato,
Phys. Lett. {\bf B302} (1993) 209.


\bibitem{YZ}
M. Yu and B. Zhang,
%{\it ``Light-Cone Gauge Quantization of String Theories on $AdS_3$ Space''},
Nucl.Phys. {\bf B551} (1999) 425,
hep-th/9812216.


\bibitem{Mizoguchi}
S. Mizoguchi,
JHEP {\bf 0004} (2000) 014, hep-th/0003053.


\bibitem{DLCQ}
L. Susskind,
hep-th/9704080;
A. Sen,
hep-th/9709220;
N. Seiberg,
Phys. Rev. Lett. {\bf 79} (1997) 3577, hep-th/9710009.

\bibitem{GRBL}
A. Giveon and M. Ro\v{c}ek,
%{\it ``Supersymmetric string vacua on $AdS_3\times {\cal N}$''},
JHEP {\bf 9904} (1999) 019, hep-th/9904024;
D. Berenstein and R.G. Leigh,
%{\it ``Spacetime supersymmetry in $AdS_3$ backgrounds''},
Phys. Lett. {\bf B458} (1999) 297,
hep-th/9904040.


\bibitem{KazS}
Y. Kazama and H. Suzuki, 
Nucl. Phys. {\bf B321} (1989) 232;
Mod. Phys. Lett. {\bf A4} (1989) 235.


\bibitem{KutS1}
D. Kutasov and N. Seiberg,
%{\it "Noncritical Superstrings"},
Phys. Lett. {\bf B251} (1990) 67;
D. Kutasov,
%{\it "Some Properties of (Non) Critical Strings"},
Lecture given at ICTP Spring School, Trieste 1991,
hep-th/9110041.

\bibitem{ES}
T. Eguchi and Y. Sugawara,
Nucl. Phys. B577 (2000) 3,
hep-th/0002100.

\bibitem{AFK}
I. Antoniadis, S. Ferrara and C. Kounnas,
Nucl. Phys. {\bf B421} (1994) 343, hep-th/9402073.

\bibitem{GK}
A. Giveon and D. Kutasov,
%{\it ``Little String Theory in a Double Scaling Limit''},
JHEP {\bf 9910} (1999) 034, 
hep-th/9909110.


\bibitem{BrF}
P. Breitenlohner and D. Freedman, 
Phys. Lett. {\bf B115} (1982) 197, 
Ann. Phys. NY {\bf 144} (1982) 249.

\bibitem{DF}
Vl. S. Dotsenko and V. A. Fateev,
Nucl. Phys. {\bf B240 [FS12]} (1984) 312.


\bibitem{KutS2}
D. Kutasov and N. Seiberg,
%{\it "Number of Degrees of Freedom, Density of States and Tachyons in
%String Theory and CFT"},
Nucl. Phys. {\bf B358} (1991) 600;
N. Seiberg,
%{\it "Notes on Quantum Liouville Theory and Quantum Gravity"},
Prog. Theor. Phys. Suppl. {\bf 102} (1990) 319.

\bibitem{Sugawara}
Y. Sugawara
%UT-859, 
Nucl. Phys. B576 (2000) 265, hep-th/9909146.

\bibitem{HS1}
K. Hosomichi and Y. Sugawara,
JHEP {\bf 9901} (1999) 013, hep-th/9812100.


\bibitem{KLL}
D. Kutasov, F. Larsen, and R. Leigh,
%{\it ``String Theory in Magnetic Monopole Backgrounds''},
Nucl.Phys. {\bf B550} (1999) 183, hep-th/9812027.


\bibitem{MS}
J. Maldacena and  A. Strominger,
%{\it ``$AdS_3$ Black Holes and a Stringy Exclusion Principle''},
JHEP {\bf 9812} (1998) 005,
hep-th/9804085.



\bibitem{MST}
R. Dijkgraaf, E. Verlinde, and H. Verlinde,
%{\it ``Matrix String Theory''},
Nucl. Phys. {\bf B500} (1997) 43,
hep-th/9703030;
L. Motl, 
%{\it ``Proposals on nonperturbative superstring
% interactions''},
hep-th/9701025;
T. Banks and N. Seiberg,
%{\it ``Strings form Matrices''}
hep-th/9702187.

\bibitem{Carlip}
S. Carlip, 
Class. Quant. Grav. {\bf 15} (1998) 3609,
hep-th/9806026.

\bibitem{ABKS}
O. Aharony, M. Berkooz, D. Kutasov and N. Seiberg,
JHEP {\bf 9810} (1998) 004,
hep-th/9808149.

\bibitem{GiveonKP}
A. Giveon, D. Kutasov and O. Pelc,
%{\it ``Holography for Non-Critical Superstrings''},
JHEP {\bf 9910} (1999) 035.
hep-th/9907178.

\bibitem{Brower}
R. Brower, Phys. Rev. {\bf D6} (1972) 1655.

\end{thebibliography}
\end{document}